\documentclass[acmsmall]{acmart}

\usepackage{amsmath}
\usepackage{mathtools}
\usepackage{amsfonts}
\usepackage{amsthm}
\usepackage[inline]{enumitem}
\usepackage{array}
\usepackage{semantic}
\usepackage{amsthm}
\usepackage{graphicx}
\usepackage{subcaption}
\usepackage{hyperref}
\usepackage[capitalise]{cleveref}
\usepackage{listings}
\usepackage{tikz}
\usepackage{xcolor}
\usepackage{accents}

\usepackage{algorithm}
\usepackage{algorithmicx}
\usepackage[noend]{algpseudocode}

\usepackage[old]{old-arrows}

\usetikzlibrary{shapes.geometric, arrows}
\theoremstyle{definition}
\newcommand{\len}[1]{\mathrm{length}~#1}

\newcommand{\tagsc}[1]{\tag{\textsc{#1}}} % Fake Small Caps tagging

\newcommand{\fvseqCustom}[3]{\bigwedge\limits^{#2}_{#1}({#3})}
\newcommand{\union}[4]{\bigcup\limits^{{#3}}_{#1={#2}}.{#4}}

\newcommand{\hsp}{\hspace{1em}}

\newcommand{\kw}[1]{\mathtt{#1}}

\newcommand{\vd}{\vdash}

\newcommand{\sembox}[1]{\hfill \normalfont \mbox{\fbox{\(#1\)}}}
\newcommand{\sempart}[2]{\textrm{\textit{#1 \sembox{#2}}}}

\newcommand{\FV}[1]{\mathcal{FV}(#1)}
\newcommand{\BV}[1]{\mathcal{BV}(#1)}

\newcommand{\lr}[1]{\left( #1 \right)}
\newcommand{\Env}{\Gamma}
\newcommand{\unify}[2]{\mathrm{unify}\lr{#1,~#2}}

\newcommand{\True}{\mathbf{true}}
\newcommand{\False}{\mathbf{false}}

\newcommand{\Ksym}{\mathcal{K}}
\newcommand{\Gsym}{\mathcal{G}}
\newcommand{\K}[1]{\Ksym\langle #1 \rangle}
\newcommand{\G}[1]{\Gsym\bigl\langle #1 \bigr\rangle}
\newcommand{\Hole}{\square}

\newcommand{\Fun}[2]{\lambda#1.\, #2}
\newcommand{\Maptt}{\texttt{map}}
\newcommand{\Map}[2]{\Maptt~#1~#2}
\newcommand{\Scantt}{\texttt{scan}}
\newcommand{\Scan}[3]{\Scantt~#1~#2~#3}
\newcommand{\Scattertt}{\texttt{scatter}}
\newcommand{\Scatter}[3]{\Scattertt~#1~#2~#3}
\newcommand{\Iotatt}{\texttt{iota}}
\newcommand{\Iota}[1]{\Iotatt~#1}
\newcommand{\Replicate}[2]{\text{replicate}~#1~#2}

\newcommand{\Iff}[3]{\kw{if}~#1~\kw{then}~#2~\kw{else}~#3}

\newcommand{\Let}[3]{\kw{let}~#1~\mbox{\texttt{=}}~#2~\kw{in}~#3}

\newcommand{\Def}[3]{\kw{def}~#1~#2~\kw{=}~#3} % top-level def

\newcommand{\ixfnNoDom}[2]{
  #1~.~#2
}
\newcommand{\forix}{
  \mathrm{f\kern-0.15emo\kern-0.06emr}
}
\newcommand{\ixfn}[2]{
  \forix~\ixfnNoDom{#1}{#2}
}
\newcommand{\iot}[2]{
  #1 < #2
}
\newcommand{\seg}[2]{
  #1 \geq #2
}
\newcommand{\cat}[2]{
  \bigcup\limits_{#1}^{#2}
}
\newcommand{\gdelim}{\bigwedge}
\newcommand{\ges}[4]{\gdelim\limits_{#1=#2}^{#3}(#4)}

\DeclareMathOperator{\nd}{\wedge}
\DeclareMathOperator{\oo}{\vee}

\newcommand{\Sum}[4]{\sum_{#1=#2}^{#3} (#4)}
\newcommand{\Rec}{\circlearrowleft}

\newcommand{\lab}[1]{{\textsc{\scriptsize (#1)}}}

\newcommand{\scalar}{\kern-0.2em\bullet\kern-0.3em}
\newcommand{\fresh}[1]{\mathrm{fresh}~#1}

\newcommand{\repSym}{\sigma}
\newcommand{\rep}[1]{\repSym(#1)}

\newcommand{\Alg}{\Delta}
\newcommand{\threeqm}{\mathop{?\kern0.1em\!\raisebox{0.3em}{?}\!\kern0.1em?}}
\newcommand{\esrc}{e^{*}}
\newcommand{\ranges}[2]{{#1}^{+#2}}

\newcommand{\query}{\overset{?}{=>}}
\newcommand{\?}{\query}

\newcommand{\RCD}{\mathrm{Rcd}}
\newcommand{\RCDimg}{\mathrm{Img}}

\newcommand{\To}[1]{\overset{#1}{\to}}

\usepackage{soul}

\definecolor{eclipseBlue}{RGB}{42,0.0,255}
\definecolor{eclipseGreen}{RGB}{63,127,95}
\definecolor{eclipsePurple}{RGB}{127,0,85}

\lstdefinelanguage{futhark}
{
  morekeywords={
    bool,
    do,
    else,
    if,
    in,
    include,
    let,
    loop,
    map,
    map2,
    map3,
    scan,
    scatter,
    then,
    type,
    val,
    while,
    with,
    module,
    def,
    entry,
    local,
    open,
    import,
    assert,
    match,
    case,
  },
  morekeywords=[2]{ % Using [2] to create a second keyword class
    FiltPartInv,
    FiltPart,
    Range,
    Inj,
    Bij,
    Mono,
  },
  basicstyle=\ttfamily\small, % Global Code Style
  sensitive=true, % Keywords are case sensitive.
  morecomment=[l]{--}, % l is for line comment.
  commentstyle=\itshape\color{eclipseGreen}, % style of comments
  morestring=[b]", % Strings are enclosed in double quotes.
  numberstyle=\scriptsize\ttfamily, % style of the line numbers
  showstringspaces=false, % lets spaces in strings appear as real spaces
  emphstyle=\ttfamily\bfseries,
  keywordstyle=\bfseries, %\textbf,         % Style for regular keywords (class 1)
  keywordstyle=[2]\color{blue},  % Style for special keywords (class 2)
  escapechar=@,
}

\begin{document}

\title{Verifying Properties of Index Arrays in a Purely-Functional Data-Parallel Language}

\author{Nikolaj Hey Hinnerskov}
\email{nhin@912134.xyz}
\orcid{0000-0001-7559-0939}
\affiliation{%
  \institution{University of Copenhagen}
  \country{Denmark}
}
\author{Robert Schenck}
\email{r@bert.lol}
\orcid{0000-0001-5848-8166}
\affiliation{%
  \institution{Vrije Universiteit Amsterdam}
  \country{Netherlands}
}
\author{Cosmin E. Oancea}
\email{cosmin.oancea@di.ku.dk}
\orcid{0000-0001-5421-6876}
\affiliation{%
  \institution{University of Copenhagen}
  \country{Denmark}
}

\begin{abstract}
  This paper presents a novel approach to automatically verify properties of
  pure data-parallel programs with non-linear indexing---expressed as pre- and
  post-conditions on functions. Programs consist of nests of second-order array
  combinators (e.g., map, scan, and scatter) and loops.
  The key idea is to represent arrays as index functions: programs are index
  function transformations over which properties are propagated and inferred.
  Our framework proves properties on index functions by distilling them into
  algebraic (in)equalities and discharging them to a Fourier-Motzkin-based
  solver.
  The framework is practical and accessible: properties are not restricted to a
  decidable logic, but instead are carefully selected to express practically
  useful guarantees that can be automatically reasoned about and inferred. These
  guarantees extend beyond program correctness and can be exploited by the
  entire compiler pipeline for optimization.
  We implement our system in the pure data-parallel language Futhark and
  demonstrate its practicality on seven applications, reporting an average
  verification time of 1 second.
  Two case studies show how eliminating dynamic
  verifications in GPU programs results in significant speedups.
\end{abstract}

%% 2012 ACM Computing Classification System (CSS) concepts
%% Generate at 'http://dl.acm.org/ccs/ccs.cfm'.
%\begin{CCSXML}
%  <ccs2012>
%  <concept>
%  <concept_id>10011007.10011006.10011008.10011009.10011012</concept_id>
%  <concept_desc>Software and its engineering~Functional languages</concept_desc>
%  <concept_significance>500</concept_significance>
%  </concept>
%  <concept>
%  <concept_id>10011007.10011006.10011008.10011009.10010175</concept_id>
%  <concept_desc>Software and its engineering~Parallel programming languages</concept_desc>
%  <concept_significance>500</concept_significance>
%  </concept>
%  </ccs2012>
%\end{CCSXML}
%
%\ccsdesc[500]{Software and its engineering~Functional languages}
%\ccsdesc[500]{Software and its engineering~Parallel programming languages}

%% Keywords
%% comma separated list
%\keywords{Pure functional language, data parallel, array programming, verification}  %% \keywords are mandatory in final camera-ready submission
% TODO

\maketitle

\section{Introduction}

\enlargethispage{\baselineskip}

A rich body of work is dedicated to
static verification of program properties. This includes
(1) theorem provers such as Rocq~\cite{bertot2013interactive} and
    Agda~\cite{bove2009brief}, which facilitate principled reasoning through
    their dependent type systems,
(2) the systems in the F* family~\cite{Fstar,swamy2023proof}, which are
    commonly aimed at verifying low-level code by, for example, combining dependent
    type, effect-based reasoning with separation logic,
(3) work on Liquid Haskell~\cite{rondon-liquid-2008,vazou-refinement-2014},
    which is facilitated by reasoning on recursive data types, such as lists,
    and allows specification of arbitrary user-defined
    properties, while enabling automated SMT-based reasoning, and
% XI_2007,DependentML-bounds-checking,xi2017applied,array-folds-logic,Qube
(4) systems that implement decidable subsets of array
    logic~\cite{DependentML-bounds-checking,Qube},
    but which are restricted to linear indexing.
Such systems are commonly aimed at sequential code and either
restrict the domain of supported computations (4),
or require expert knowledge (e.g., the programmer
must explicitly compose the proofs checked by the system).

% Maybe something about SMT solvers being computationally expensive?
%

This paper presents compiler analyses for inferring
and verifying properties of {\em integral arrays}---e.g., ranges,
monotonicity, injectivity, % equivalences
bijectivity, filtering, partitioning---applicable to {\em functional
data-parallel languages}, such as 
Futhark~\cite{henriksen2017futhark,stl-climate,app-kd-tree},
Accelerate~\cite{10.1145/1926354.1926358,accelerate-nested,accelerate-scans}, 
Lift~\cite{Lift-orig,Lift,steuwer2022rise},~DaCe~\cite{dace-stateful,dace-climate,dace-quantum-transport}.

Specific to this domain, the computation is {\em separated} (fissed) into
bulk-parallel array operations---such as map, scan (prefix sum), and scatter
(irregular write)---called \emph{second-order array combinators},
together~with~loops. Eventually, index arrays computed in this way
are serve as the indirect indices of irregular read (gather) or
write (scatter) operations.
%
%%This presents challenges to analysis in comparison to the sequential
%%setting, where natural expression fuses the computation,
%%(e.g., into folds), which facilitates tracking the target property
%%each step of the way.
%
%Many bulk-parallel operations manipulate integral arrays that
%eventually serve as the indirect indices of irregular read
%(gather) or write (scatter) operations.
%
Index arrays are analytically convenient because (1) they are
typically manipulated in simpler ways than general computation,
and (2) they directly inform properties of general arrays, such
as injectivity, filtering, and partitioning.

The programming style is both a challenge and an opportunity:
On the one hand, the natural expression of sequential computation
is in fused form, (e.g., folds), which facilitates easier tracking
of the target property at each step of the way. On the other hand,
language purity and the semantics of second-order combinators
allow to lift the level of abstraction at which the compiler reasons.

\enlargethispage{\baselineskip}

Our solution is motivated by the evolution of scheduling
languages~\cite{sched-lang,Halide}, % ,venkat:pldi15
which demonstrates that the coupling of language specalization (and its compiler
repertoire) with human expertise has been essential to unlocking high
performance.
In the same spirit, our system supports (automates) a predefined, small
but powerful set of properties, which are easy to understand and use by %understandable by the
the {\em domain expert} (non-expert programmer).  %\rs{See *2 below}
This allows the compiler to exploit the algebra of
properties to derive new properties at a high level, and also to use the proven
properties to further optimize the program. Possibilities here include static
verification of bounds checking and safety of scatter, which we
demonstrate (\cref{sec:eval}) to have high impact on GPU execution. On the other
hand, our system does not allow the user to specify new properties, and is
intended to be neither satisfiable nor decidable---we aim at practical compilation
time without restricting the language.

Our framework is aimed at verifying Futhark programs and is implemented as a compiler
pass that is structured into three logical components: (1) an analysis
\textsc{InfIxf}, presented in \cref{sec:infer-ixfn}, that infers an
index function representation of arrays, which uses guarded expressions
(polynomials defined by cases) to represent the values at each array index and
supports jagged arrays whose segments may be empty, and (2) a property manager
\textsc{PM} (\cref{sec:prove-props}), that verifies properties of index
functions by breaking them into a sufficient set of low-level queries, which are
sent to (3) the query solver \textsc{QS} (\cref{sec:query-solver}), which uses a
Fourier-Motzkin~\cite{Fourier-orig,Williams01111986} adaptation algorithm
that relies on an algebra of simplifications aimed at sums of array slices.

These components are connected by means of the supported
array properties, e.g., the inference rule of scatter uses
monotonicity and bijectivity properties to produce meaningful
index functions that enable expression of jagged arrays and
derivation of filtering/partitioning properties. The property
manager can derive properties at a high level, in the absence
of an index function, e.g., a filtering of a monotonic or
injective array remains monotonic or injective. Finally, the
query solver uses range, injectivity and monotonicity
properties, and answers the queries of
\textsc{InfIxf}~and~\textsc{PM}.

\Cref{sec:eval} presents an evaluation of verifying properties on seven
challenging data-parallel applications that use non-linear indexing, including
the maximal matching graph algorithm, three-way partitioning, segmented
filtering and segmented two-way partitioning.\footnote{
  Segmented two-way
  partitioning is the flat-parallel code~\cite{Blelloch-Nesl} that
  filters/partitions each subarray of an (irregular) jagged array
  in flat form, according to predicates that depend on the segment number.
}

The principle contributions of this work are:
\begin{enumerate}
\item To our knowledge, this is the first solution addressing verification of
  array properties---including monotonicity, bijectivity,
  filtering, partitioning---in a purely-functional data-parallel context with
  non-linear indexing (produced by scatter, gather, scan).
\item An architectural design that supports verification of a predefined set of
  properties that are accessible to non-expert programmers, and reveal to the
  compiler a
  compositional algebra that facilitates high-level reasoning, automation of
  inference and further code optimizations.

\item An evaluation that reports successful verification of
           seven challenging benchmarks that include graph algorithms
           and flattened irregular parallelism (segmented filter/partitioning).
%           and flattened code~\cite{Blelloch-Nesl} corresponding to
%           the batch application of filter and two-way partitioning.
\end{enumerate}

\newpage

%%%%%%%%%%%%%%%%%%%%%%%%%%%%%%%%%%%%%%%%%%
%%% Bird's Eye View
%%%%%%%%%%%%%%%%%%%%%%%%%%%%%%%%%%%%%%%%%%
\section{Languages, Array Properties and Bird's Eye View of Main Components}
\label{sec:birds-eye-view}

We describe our system as a static analysis
on user-written source code.
Accordingly, we begin by introducing the source language
and two running examples in \cref{subsec:src-lang-egs}.
\Cref{subsec:ixfn-rep-bev} presents an internal language used by
our analysis and the bird's eye view of the architecture. Finally,
\cref{subsec:array-props} specifies the array properties supported
by our system.

\subsection{The Source Language of the Analysis and Running Examples}
\label{subsec:src-lang-egs}

% \textcolor{gray}{[\gdelim (\neg c => \infty)

\begin{figure}
\begin{scriptsize}
  \begin{minipage}[t]{0.46\linewidth}
\[
\begin{array}{lclr}
x,F   & ::=  & $\dots$    & \text{Variables}
\medskip\\
\tau^b   & ::=  & \kw{i64} \ | \ \kw{f64} \ | \ \kw{bool} \ | \ \ldots
                                       & \text{Base Types}\\
\tau^a   & ::=  & \tau^b \hsp | \hsp [e^{*}]\tau^b
                                       & \text{Array Type} \\
\tau     & ::=  & (..,\tau^a,..)       & \text{Tuple Type}
\medskip\\
Pre   & ::=  & e^{*}                   & \text{Pre Cond.}\\
Pst   & ::=  & \lambda x.~e^{*}        & \text{Post Cond.}
\medskip\\
Fun  & ::=  & \Def{F}{\overline{(x : \tau \textcolor{gray}{~[|~Pre]})}
                                       & \text{Function Def.}
\\
      &      & \hsp\hsp :~ \tau~\textcolor{gray}{~[|~Pst]}} {\esrc}  &
\medskip\\
op     & ::=  & + ~\mid~ - ~\mid~ *    & \text{Binary Op.} \\
       & \mid & > ~\mid~ \geq ~\mid~ < ~\mid~ \leq ~\mid~ = ~\mid~ \neq
\medskip\\
v     & ::=  & x \hsp | \hsp k               & \text{Var./Const.} %\quad (k \in \mathbb{Z})
\end{array}
\]
  \end{minipage}
  \begin{minipage}[t]{0.53\linewidth}
\[
\begin{array}{lclr}
e^0   & ::=  & v                                          & \text{Var. / Const.} \\
      & \mid & v_1~op~v_2                                 & \text{Operation} \\
      & \mid & x_{array}[x_{index}]                       & \text{Array index} \\
      & \mid & \Iota{x}                                   & [0,\ldots,x-1] \\
      & \mid & \Replicate{x_n}{x}                         & [x,\ldots,x]\\
      & \mid & \Iff{x}{\esrc}{\esrc}                      & \text{Conditional} \\
%
%      & \mid & \WhileL{(\overline{x_v \textcolor{gray}{[:\tau~|~Pre]}}=\overline{x_0})}{x_c}{\esrc}
%                                                          & \text{While Loop}\\
      & \mid & \kw{loop}~(\overline{x_{var} \textcolor{gray}{~[:\tau~|~Pre]}}^q=\overline{x_{init}}^q)
                                                          & \text{While Loop}\\
      &      & \hsp \kw{while}~x_c \ \kw{do} \ \esrc_{body} & \\
      & \mid & \Map{(\Fun{\overline{x_{elm}}^q}{\esrc})}{\overline{x_{array}}^q}
                                                          & \text{Map} \\
      & \mid & \Scan{(\Fun{ \overline{x_1}^q \ \overline{x_2}^q}{\esrc})}{ \ \overline{k}^q \ }{\overline{x_{array}}^q}
                                                          & \text{Scan} \\
      & \mid & \Scatter{x_{dest}}{x_{inds}}{x_{vals}}     & \text{Scatter} \\
      & \mid & F \ \overline{x}                           & \text{Apply Fun.} \medskip\\
%      & \mid & (..,x,..)                                  & \text{Tuple Exp.}\medskip\\
%
\esrc & ::=  & \Let{(..,x,..)}{e^0}{\esrc}
                    \hsp | \hsp (..,x,..)                 & \text{Bindings}
\end{array}
\]
  \end{minipage}\smallskip

\textbf{Notation:} $\overline{o}^q$ denotes a sequence of $q$ objects of
some kind, separated by white space or by comma, as dictated by the context.
\end{scriptsize}
\vspace{-2ex}
\caption{\label{fig:source-grammar}
  Grammar for source language expressions ($\esrc$) and function declarations ($Fun$).
}
\end{figure}

The source language ($e^{*}$), with syntax shown in \cref{fig:source-grammar},
is a standard purely functional, first-order expression language, which has
been augmented with
second-order array combinators, such as $\kw{map}$ and $\kw{scan}$,
and enforces a structure-of-arrays (SoA) layout.
For presentation purposes, the language is in A-normal form~\cite{A-Normal-Form},
%
% For simplicity of exposition, the source language of the analysis,
% whose syntax is shown in \cref{fig:source-grammar}, uses a
% standard, purely functional, first-order expression language ($e^{*}$)
% in A-normal form~\cite{A-Normal-Form}, but which is augmented with
% second-order array combinators, such as $\kw{map}$ and $\kw{scan}$,
% and enforces a structure-of-array (SoA) layout.
%
% monomorphic, statically typed, strictly evaluated
%For simplicity of exposition, the presentation uses a standard,
%first-order, purely-functional array language in A-normal
%form~\cite{A-Normal-Form}, but which is augmented with
%second-order array combinators such as map and scan and which
%enforces a structure-of-array (SoA) layout.
%
meaning (1) let bindings can be
seen as a block of statements followed by a sequence of result
variables, and (2) if conditions, loop initializers and
function operands are variables.
The types conform with the SoA layout, for example, $([n]\kw{i64},~[n]\kw{bool})$
is a valid tuple type of integer and boolean arrays,
both of length $n$, but the type $[n](\kw{i64},~\kw{bool})$ is invalid,
because it uses an array-of-structures (AoS) layout.

Function declaration allows annotating argument types with preconditions
(boolean expressions) and the result type with a postcondition
(lambda function from the result type to booleans).
%
%The order of the arguments and results is important: in the case of
%a tuple type, each property (of the conjunction) is attributed to
%the latest declared argument/result that is referenced in that property,
%and similarly, pre conditions cannot refer to successor arguments.

The language has array constructors
$\kw{iota}~n$, which produces the array $[0,\ldots, n-1]$,
and $\kw{replicate}~n~x$, which produces a length-$n$ array containing $x$ at each index.
As well as if and while loops, which are equivalent to % semantically
tail-recursive functions:
the $q$ loop parameters $\overline{x_{var}}^q$ are initialized with
the values of variables $\overline{x_{init}}^q$ and are bound to the
result of the loop-body expression $e^{*}_{body}$ for the remaining
iterations.
Map and inclusive scan (prefix sum) have standard types and semantics:\vspace{-2ex}

\[
\begin{array}{lll}
\kw{map} \ : \ (\alpha \rightarrow \beta) \rightarrow [n]\alpha \rightarrow [n]\beta &
\kw{scan} \ : \ (\alpha \rightarrow \alpha \rightarrow \alpha)
  \rightarrow \alpha \rightarrow [n]\alpha \rightarrow [n]\alpha\\
\kw{map} \ f \ [x_0,\ldots,x_{n-1}] = [f~x_0,\ldots, f~x_{n-1}] &
\kw{scan} \ \odot \ ne_\odot \ [x_0,\ldots,x_{n-1}] = [x_0,\ldots,x_0 \odot\ldots\odot x_{n-1}]
\end{array}
\]\vspace{-1ex}

\noindent But they adhere to the SoA layout: their lambda functions determine an
arbitrary number of results,
and they accept a matching number of arguments; % Naturally, % arrays
$\odot$ must be associative with neutral element $ne_\odot$.

Finally, $\kw{scatter}$ is a bulk-write operator of type
$[n]\alpha \rightarrow [m]i64 \rightarrow [m]\alpha \rightarrow [n]\alpha$
and imperative semantics:
$\kw{scatter}~dst~is~vs \ \equiv \ \kw{for}~ i=0\ldots m-1 \
\{ \ \kw{if} ~ (0 \leq is[i] < m) \ dst[is[i]] ~\text{:=}~ vs[i] \ \}$. That is,
$\kw{scatter}$ updates $dst$ in place at indices $is$ with the
corresponding values from $vs$, but ignores updates to indices
that are out of bounds in $dst$.
Its pure semantics and $O(m)$ work asymptotic are ensured by a
type checking technique that builds on uniqueness
types~\cite{henriksen2017futhark}. To match the imperative, deterministic
semantics, $\kw{scatter}$ must be idempotent. We can ensure this
by requiring any duplicate indices in $is$ to correspond to equal values in $vs$:
\begin{center}
$
% 0\leq i_{1,2} < m \wedge 0 \leq is[i_{1,2}] < n \wedge is[i_1] = is[i_2] \Rightarrow vs[i_1] = vs[i_2].
0\leq i_1 < m \wedge 0\leq i_2 < m \wedge 0 \leq is[i_1] < n \wedge 0 \leq is[i_2] < n \wedge is[i_1] = is[i_2] \Rightarrow vs[i_1] = vs[i_2].
$
\end{center}

\noindent However, this is not verified in Futhark, neither statically nor dynamically;
our work enables static verification of scatter and of array indexing
(mostly verified by dynamic assertions), among others.\smallskip % currently

\textbf{Our implementation} uses Futhark's source
language~\cite{henriksen2017futhark,futhark-ppopp}, which is neither in A-normal
form nor uses an SoA layout. For clarity, we elide the details of the
transformation into this form, which is automatically performed in later
compilation stages~\cite{henriksen:phdthesis}.  In particular, we support
passing predicates as arguments to functions to enable general expression of
filtering/partitioning~properties.

Futhark has support for size-dependent types~\cite{sized-types,bailly2023shape},
but size equality is purely syntactical: $[n+m]\tau \not= [m+n]\tau$.
Our work naturally extends the expressible size dependencies
using pre- and postconditions.
%e.g., by supporting expressions containing sums of array slices
%
The modifications needed to accommodate pre- and postconditions are standard:
preconditions are verified and postconditions are assumed at call sites,
and vice versa for function declarations.
A property assigned to a loop parameter is verified on the loop
initializer; then the property is assumed on the
parameter and is verified on the corresponding loop-body result.
\smallskip

\begin{figure}
\begin{scriptsize}
  \begin{minipage}[t]{0.40\linewidth}
\begin{lstlisting}[language=futhark,basicstyle=\scriptsize,numbers=left,mathescape=true]
let sum [n] (xs: [n]i64) =
  if n > 0 then (scan (+) 0 xs)[n-1] else 0

def partition2 [n] (p: f64 -> bool) (xs: [n]f64)
    : (i64, [n]f64) | \ (m,ys) ->
       m == sum ( map i64.bool (map p xs) )            @\label{line:postcond-m}@
       && FiltPart ys xs (\_->true) (\i-> p xs[i]) =
  let cs = map (\x -> p x) xs
  let flagsT = map (\c -> if c then 1 else 0) cs
  let flagsF = map (\b -> 1 - b) flagsT
  let indicesT = scan (+) 0 flagsT
  let num_true = if n > 0 then indicesT[n-1] else 0
  let tmp      = scan (+) 0 flagsF
  let indicesF = map (\t -> t + num_true) tmp
  let indices  = map3 (\c t f -> if c then t-1 else f-1)  @\label{line:inds-of-scatter}@
                    cs indicesT indicesF
  let zeros = replicate n 0
  let ys = scatter zeros indices xs @\label{line:scatter-inds}@
  in (num_true, ys)
\end{lstlisting}
%
%   let zeros = map (\_ -> 0) (iota n)
%
  \end{minipage}
  \begin{minipage}[t]{0.12\linewidth}
\begin{lstlisting}[language=futhark,basicstyle=\scriptsize,mathescape=true]



$\Longleftarrow$ Demo
p = \ x -> x < 5,
xs= [5,4,2,8,7,3]
n = 6
[F, T, T, F, F,T]
[0, 1, 1, 0, 0, 1]
[1, 0, 0, 1, 1, 0]
[0, 1, 2, 2, 2, 3]
3
[1, 1, 1, 2, 3, 3]
[4, 4, 4, 5, 6, 6]
[3, 0, 1, 4, 5, 2]

[0, 0, 0, 0, 0, 0]
[4, 2, 3, 5, 8, 7]
\end{lstlisting}
  \end{minipage}
  \begin{minipage}[t]{0.46\linewidth}
  \begin{lstlisting}[language=futhark,basicstyle=\scriptsize,numbers=right,mathescape=true]
def sgmSum [n] (flags: [n]bool) (xs: [n]i64) : [n]i64 =                         @\label{line:sgm-scan-def}@
  let (_, ys) = scan (\ f1 v1 f2 v2 -> let f = fx || fy
                     let v = if fy then vy else vy + vx
                     in (f, v)     ) false 0i64 flags xs in ys
def mkSgmDescr [m] (shape: [m]i64 | Range shape (0,$\infty$))
    (xs: [m]i64) : []t | (\flags -> length flags == sum shape) =
  let rot = map (\i -> if i==0 then 0 else shape[i-1]) (iota m)  @\label{line:rotate}@
  let scn = scan (+) 0i64 shp_rot
  let ind = map2 (\s i -> if s<=0 then -1 else i) shape shp_scn  @\label{line:neg-index}@
  let len = if m > 0 then shp_scn[m-1] + shape[m-1] else 0
  let res = scatter (replicate len 0) ind xs    in res
  -- Example: shape=[0,2,1,0,3] & xs=[1,2,3,4,5] $\Rightarrow$ res=[2,0,3,5,0,0]
def mkII [m] (shape: [m]i64 | Range shape (0,$\infty$))
      : []i64 | (\II -> length II == sum shape) =
  let beg_vs = map (\i -> i + 1) (iota m) $\phantom{lala}$ -- [1,2,3,4,5]
  let sct_vs1= mkSgmDescr shape beg_vs   $\phantom{lal}$ -- [2,0, 3, 5,0,0]     @\label{line:call-mkFlag}@
  let sct_vs = map (\v -> if v == 0 then 0 else v-1) scat_vs1                   @\label{line:beg-sgm-ind}@
  let flags = map (\f -> f > 0) scat_vs  $\phantom{lalala}$ -- [T,F, T, T,F,F]  @\label{line:flag-array}@
  let II    = sgmSum flags scat_vs in II $\phantom{lalalaa}$ -- [1,1, 2, 4,4,4] @\label{line:sgm-scan-call}@
\end{lstlisting}
  \end{minipage}
\end{scriptsize}
\vspace{-2ex}
\caption{\label{fig:running-egs}
  Running Examples: two-way partitioning (left) with demo (center) \& building flag and II arrays (right).
}
\end{figure}

\textbf{Running Examples.}
Figure~\ref{fig:running-egs} shows two code examples that are
illustrative for the functional data-parallel programming style, where
the computation is separated (fissed) into a sequence of bulk-parallel
operations. Many such operations manipulate integral arrays that are
eventually used for indirect indexing in gather and scatter operations.
For example, the code on the left implements a two-way partitioning
of an array $xs$ based on a predicate $p$: the predicate is first mapped
across the elements of $xs$ and the integral result ($flagsT$) is scanned,
resulting in array $indicesT$ that holds the indices at which
the elements that succeed should be scattered plus one. The failing indices
are treated similarly, resulting in $indicesF$. The final indices
are put together by the $\kw{map}$ operation at line~\ref{line:inds-of-scatter},
and, finally, the partitioned arrays is computed by the scatter at
line~\ref{line:inds-of-scatter}. This makes it more challenging to
verify the partitioning property than in sequential languages, where
the computation is, e.g., aggressively fused into a fold, whose
accumulator maintains separate lists of succeeding and failing elements,
thus allowing to track the property across each statement.
%
%($\kw{fold}$ however does not have parallel semantics).
%
The post conditions of {\tt partition2} are that the split point $m$ is equal to
the number of elements that succeed under $p$ and that the result is a
partitioning of $xs$ (as detailed in \cref{subsec:array-props}). % subsec:bev-props

The right hand side of~\cref{fig:running-egs} shows helper functions that are
used in flattened/segmented computations: $mkSgmDescr$ takes as arguments the
shape of a jagged array---which is allowed to have empty segments (the
precondition only requires non-negative elements)---and an array $xs$ of
matching length. It returns a flat array of length equal to the sum of the
shapes (see the postcondition), such that each non-empty segment starts with the
corresponding element of $xs$ and the rest of the segment is zeroed.  For
example, if $shape=[0,2,1,0,3]$ and $xs=[1,2,3,4,5]$ then
$mkSgmDescr~shape~xs~=~[2,0,3,5,0,0]$.  Note that empty segments pass
negative indices (line~\ref{line:neg-index}) to scatter, which are ignored
(otherwise WAW races would violate its deterministic semantics).

Similarly, $mkII$ in \cref{fig:running-egs} takes as argument a shape array and
produces a flat array of that shape in which each element is assigned the index
of the segment in which it resides. For example, if $shape = [0,2,1,0,3]$ then
$mkII~shape = [1,1,~2,~4,4,4]$. We will name this array $II$ henceforth.
The~implementation~uses $mkSgmDescr$ (line~\ref{line:call-mkFlag}) to inscribe
the index of each non-empty segment at the start of the segment
(line~\ref{line:beg-sgm-ind}), i.e., $[1,0,~2,~4,0,0]$, and then propagates the
start element throughout the segment by using a segmented scan~\cite{segScan}
(line~\ref{line:sgm-scan-call}), which is implemented as a scan with a lifted
operator % classically
(line~\ref{line:sgm-scan-def}).  Note that there are no standard properties
that accurately summarize the results of $mkII$ and $mkSgmDescr$, albeit their
content can be easily described as index functions.

%create a flag array (line~\ref{line:flag-array})

%%%%%%%%%%%%%%%%%%%%%%%%%%%%%%%%%%%%%%%%%%%%%%%
%%% IxFn Representation and Bird's Eye View
%%%%%%%%%%%%%%%%%%%%%%%%%%%%%%%%%%%%%%%%%%%%%%%

\subsection{Index-Function Representation and Bird's Eye View of Architecture}
\label{subsec:ixfn-rep-bev}

\begin{figure}
\begin{scriptsize}
  \begin{minipage}[t]{0.33\linewidth}
    \[
    \begin{array}{l}
    \begin{array}{cclr}
    c  & ::=  & x       & \text{Variable}\\
       & \mid & n       & \text{Integer}\\
       % & \mid & 2^{e}& \text{Pow. Two}\\
       & \mid & x[e] & \text{Indexing}\\
       & \mid & \Sum{x}{e}{e}{c}  & \text{Sum}\\
       & \mid & x^{-1} & x[x^{-1}[i]] = i\\
       & \mid & \infty  & \text{Special: $\infty \not\in \mathbb{Z}$}\\
       & \mid & \Rec  & \text{Recurrence}\\
       % & \mid & \neg c & \text{Logical neg.}\\
       & \mid & e \odot e & \text{Comparison}\\
       & \mid & \neg c    & \text{Logicals}\\
       & \mid & c \land c ~\mid~ c \lor c
       % & \mid & c \diamond c ~\mid~ \neg c \hspace{-2.5em}& \text{Logicals}
       % \smallskip
    \end{array}
    % \\
    % \begin{array}{ccl}
    % \diamond & ::= & \land ~\mid~ \lor
    % \end{array}
    \end{array}
    \]
  \end{minipage}\begin{minipage}[t]{0.44\linewidth}
    \[
    \begin{array}{l}
    \begin{array}{cclr}
      t   & ::= & c ~\mid~ c \cdot t & \text{Term}\smallskip\\
      e   & ::= & t ~\mid~ t + e & \text{Polynomial}\smallskip\\
      g  & ::= & c => e ~\mid~ g \gdelim g & \text{Guarded expr.}\\
      D  & ::= & \forix~x~<~e ~\mid~ \union{k}{0}{e}{x \geq e} & \text{Lin/Sgm dom.}
    \end{array}\\
    \begin{array}{ccl}
    ixfn & ::=    & D.~g \hfill \text{Index function}\\
         & ~\mid~ & (\text{for}~x~<~e.g) \ \cup \ (\union{k}{1}{e}{x \geq e\propto e}.~~g)
    \end{array}\\
    \begin{array}{ccl}
    \odot    & ::= & < ~\mid~ \leq ~\mid~ > ~\mid~ \geq ~\mid~ = ~\mid~ \neq
    \end{array}
    \end{array}
    \]
  \end{minipage}\begin{minipage}[t]{0.20\linewidth}
    \[
    \begin{array}{l}
    \textbf{Notation:}\\
    i,j,k \hfill\text{integral iterators}\\
    v,w \hfill \text{enumeration bounds}\\
    h,l \hfill\text{ct-range iterators}\\
    x,y,z \hfill\text{array variables}\\
    e\{x\mapsto e_x\} \hfill \text{substitutes $x$}\\
    \hfill \text{for}~e_x~\text{in}~e
    \\
    \ges{h}{1}{v}{c_h => e_h} \quad\text{denotes}\\
    \hsp \hfill c_1 => e_1 \gdelim \dots \gdelim c_v => e_v
    \end{array}
    \]
  \end{minipage}

\end{scriptsize}
\vspace{-2ex}
\caption{Grammar for symbols ($c$), polynomials ($e$), guarded expressions ($g$), and index functions ($ixfn$).}
\label{fig:internal-grammar}
\end{figure}

Figure~\ref{fig:internal-grammar} presents the grammar for the internal languages.
A symbol ($c$) can be a variable name,
an integer constant,
%a power of two (to accommodate FFT and the like),
array indexing,
a sum of symbols,
an inverse array (i.e., $x^{-1}[x[i]] = x[x^{-1}[i]] = i$;
see~\cref{sec:scatter}),
a special $\infty$ symbol used for pattern matching,
a recurrence (needed transiently to represent scans---see \cref{sec:scan}),
a comparison,
or a Boolean operation.
All constants are integers; in Boolean operations 0 is considered
false, and 1 is considered true.\footnote{
 This treatment lets us sum over boolean symbols, for example,
 a sum over the number of true guards in an index function.
}
% so consider Boolean logic to
% be encoded using integers.
%
We normalize addition and multiplication on symbols into
multivariate polynomials ($e$)---for instance,
$7 + 2 \cdot x^2\cdot y^3 +3\cdot x \cdot y \cdot z^2$---in which the order
of symbols in a term and the order of terms in a polynomial are syntactically
and semantically irrelevant.\footnote{
  A term is implemented as a multiset and a polynomial as a mapping that binds each
  term to its integer coefficient.
}
If symbols are restricted to variable names, the representation is strongly
normalizing: two expressions are semantically equivalent {\em iff} they are
syntactically equal. As it is, it is not (e.g., $\Sum{i}{0}{5}{x[i]}$ is equivalent
to $\Sum{i}{0}{4}{x[i]} + x[5]$), albeit our simplification engine
attempts to keep them as normalized as possible (see \cref{subsec:simplify}).
%
% The order in which symbols appear in polynomials in our text is
% syntactically irrelevant.

%

Index functions are represented by an iteration domain ($D$) followed by a finite set
of guarded expressions ($g$) delimited by $\gdelim$. The guards partition
the domain, such that the expression whose guard is true (non-zero) provides
the value at each index of the domain. For example, variable {\tt rot} at line~\ref{line:rotate}
in~\cref{fig:running-egs} has index function
$\text{for}~i<m.~(i = 0 \Rightarrow 0) \gdelim (i\not=0 \Rightarrow \text{shape}[i-1])$.
Consequently, we maintain that guards are mutually exclusive and their disjunction
is a tautology.
We treat scalars as arrays of length one
and in guards we may write
$\kw{true}$ and $\kw{false}$ for 1 and 0, for clarity.
% Keeping the only value type as integer allows us to
% straightforwardly sum over guards.

% We convert to DNF at query (explained later).
% Guards ($c$) are maintained in disjunctive normal form (DNF) since they
% are often used as premises for queries and DNF allows an accurate split into
% sub-queries that~have~conjunctions~as~premises:\\
% $C_1 \vee C_2 \Rightarrow Q \ \equiv \ (\neg~(C_1 \vee C_2)) \vee Q \ \equiv \ (\neg C_1 \vee Q) \ \wedge \ (\neg C_2 \vee Q) \ \equiv \ (C_1 \Rightarrow Q) \ \wedge \ (C_2 \Rightarrow Q)$
%

%

\subsubsection*{Domains}
\label{sec:domains}
We define three kinds of domains. A {\em linear} domain has form
$for~i<e^n$ and denotes that iterator $i$ takes values in $0 \ldots e^n-1$.
A \emph{segmented} domain $\bigcup_{k=0}^{e^m} j \geq e_k$ requires that $e_k$ is
(non-strictly) monotonic in $k$ and $e_k\{k\mapsto 0\} = 0$, and denotes
the union of integral intervals $[e_k, e_{k+1})$,
where $k=0\ldots e^m$ and $e_{k+1} = e_k\{k\mapsto k+1\}$.
The representation has \textbf{\em two key properties}:
\begin{itemize}
\item[1] Permitting empty intervals is essential to expressing
      jagged arrays such as the $II$ result of $mkII$ in~\cref{fig:running-egs}
      as:
$
\bigcup_{k=0}^{m-1} ~j \geq \sum_{k'=0}^{k-1}(shape[k']).~ \True \Rightarrow k
$
or the result of $mkSgmDescr$ as
$
\bigcup_{k=0}^{m-1} ~j \geq \sum_{k'=0}^{k-1} \text{shape}[k'].~(j=\sum_{k'=0}^{k-1} \text{shape}[k'] \Rightarrow xs[k]) \ \bigwedge \ (j>\sum_{k'=0}^{k-1} \text{shape}[k'] \Rightarrow 0)\smallskip
$
\item[2] A segmented index function $\bigcup_{k=0}^{e^m} i \geq e_k.~g$ can always be
      translated to a linear domain by using the $II$ array, which records
      at some index $i$ the segment in which $i$ resides, i.e., $k= II[i]$. The
      linear translation is thus $for ~i<e_k\{k\mapsto e^m+1\}.~g\{k\mapsto II[i]\}$.
\end{itemize}
%to express the index function of the result {\tt res} of
%$mkSgmDescr$ in~\cref{fig:running-egs} as:

%\rs{This is confusing, since we started with ``Three kinds of domains'' and now it's talking about
%  kinds of index functions. Also, I don't think $\propto$ is properly introduced.}
%The last kind of index function
The third kind models interval $[0, e^n)$ as the union of a linear
domain with a segmented one:
$(\text{for}~i~<~e_0.~g^{lin}) \ \cup \ (\bigcup_{k=1}^{e^m}.~j \geq e_k\propto e^n.~g^{sgm})$
%covers the domain $0 \ldots e^n-1$ with a
denotes a first interval $[0, e_0)$ in linear form and
$e^m$ intervals $[e_k, e_{k+1})$ in segmented form.
Denoting by $e_{m} = e_k\{k\mapsto e^m\}$, the last interval is
$[e_m,~e^n)$, and it must hold that $e_m \leq e^n$ and
$e_0 = e_k\{k\mapsto 0\}$. This kind is motivated by the case
of  scattering at the positions of $e^m$ positive and monotonic indices,
which naturally produces $e^m+1$ intervals.\medskip

\vspace{-0.2em}
\subsubsection*{Extensions}
Potential useful extensions of index functions
include allowing arbitrary union of linear and segmented domains that
can be achieved with the grammar extension:
\begin{center}
$
\hsp\hsp D ::= \text{for} <~e ~\mid~ \cup_{k=e}^{e} \geq e \hsp\hsp S ~::=~ D.g~\mid~S \cup S \hsp\hsp ixfn = \text{Dim}~x < e.~S
$
\end{center}
\noindent which also allows a general representation of multi-dimensional
array by nesting domains; our implementation supports the simple case of
2D arrays having linear domains on each dimension.
Finally, the index function may be lifted to support guards that are
invariant to the domain iterators, i.e.,
$ixfn^{gen} ::= (c \Rightarrow ixfn) \ \mid \ ixfn^{gen} \wedge ixfn^{gen}$,
which would improve the treatment of if expressions.
The treatment of these extensions is tedious,
but relatively straightforward, i.e., they can be reasoned by combining
the treatments of linear and segmented domains.
\medskip

\begin{figure}
\begin{scriptsize}
  \begin{minipage}[t]{0.3\linewidth}
  \hspace{3ex}\textsc{Infer}~\textsc{Index}~\textsc{Functions}~(\textsc{InfIxf})\hspace{3ex}\hfill\\
  \fbox{
  $
  \begin{array}{l}
  \Env \ \ \textsc{maps variable names to IxFn}\\
  \\
  \bullet \ \textsc{Static Analysis for Inferring}\\
  \hsp \textsc{Index Functions (IxFn) from}\\
  \hsp \textsc{Source Language}\\
  \\
  \bullet \ \textsc{Inference Rules Use the}\\
  \hsp \textsc{Properties Recorded in} \ \Alg
  \end{array}
  $
  }
  \end{minipage}\begin{minipage}[t]{0.04\linewidth}
  \vspace{10ex}
  $\longrightarrow$\\
  \vspace{5ex}\\
  $\longleftarrow$\\
  \end{minipage}\begin{minipage}[t]{0.35\linewidth}
  \hspace{9ex}\textsc{Property}~\textsc{Manager}~(\textsc{PM})\hspace{3ex}\hfill\\
  \fbox{
  $
  \begin{array}{l}
  \Alg \ \ \textsc{maps variable names to properties}\hfill\\
  \\
  \bullet \ \textsc{Verifies Properties of IxFns by}\\
  \hsp \textsc{Translating them to a Set of}\\
  \hsp \textsc{Sufficient-Condition Queries}\\
  \\
  \bullet \ \textsc{Verifies/Infers Properties at a}\\
  \hsp \textsc{High Level, e.g., on Loop Results}
  \end{array}
  $
  }
  \end{minipage}\begin{minipage}[t]{0.04\linewidth}
  \vspace{3ex}
  $\longleftarrow$\\
  \vspace{3ex}\\
  $\longrightarrow$\\
  \vspace{4ex}\\
  %$\circlearrowleft$
  \includegraphics[width=0.7\linewidth]{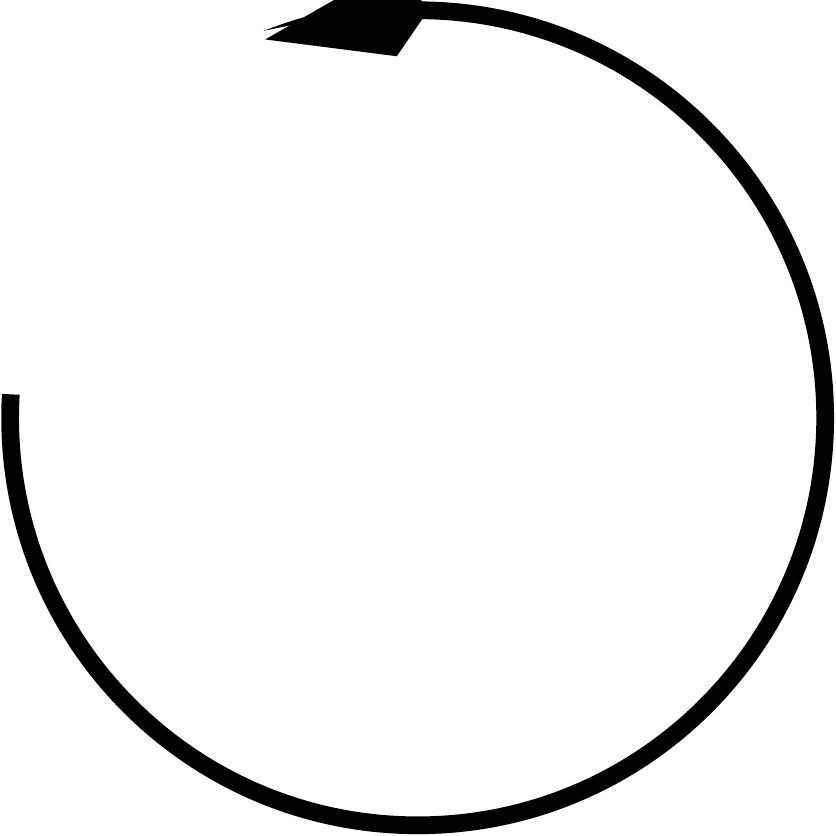}\hfill
  \end{minipage}\begin{minipage}[t]{0.3\linewidth}
  \textsc{Query Solver (QS)}\\
  \fbox{
  $
  \begin{array}{l}
  \textsc{Uses the Symbol Tables in} \ \Alg,\hfill\\
  \ \ \textsc{e.g., Ranges, Equivalences,}\\
  \ \ \textsc{Monotonicity, Injectivity,}\\
  \ \ \textsc{to Solve an Inequality or}\\
  \ \ \textsc{an Equation in the Poly}\\
  \ \ \textsc{Representation.}
  \\
  \\
  \\
  \end{array}
  $
  }
  \end{minipage}
\end{scriptsize}
\vspace{-2ex}
\caption{Bird's Eye View of the Three Logical Components of the Framework, which are deeply connected.}
\label{fig:bev-log-comps}
\end{figure}

\vspace{-0.2em}
\subsubsection*{Architectural Components}
The verification analysis is performed
during one traversal of the source program, but is split into three
connected logical components, summarized in~\cref{fig:bev-log-comps}.

\textsc{InfIxf} (left) corresponds to static analysis of function declarations
in AoS, A-Normal form that infers index-function representations for the scalar and
array variables of integral base types (and for other types in special cases).
These are recorded in the symbol table $\Env$.  \textsc{InfIxf}'s inference
rules commonly require verification of simple (in)equalities, which are
delegated to the query solver (\textsc{QS}), while challenging constructs such
as $\kw{scatter}$, also require %(verification of) properties such as
monotonicity or bijectivity.

The second component, the property manager (\textsc{PM}), handles the recording
and verification of properties, such as ranges, equivalences, monotonicity,
injectivity, bijectivity, filtering/partitioning. Properties are recorded
into corresponding symbol tables, aggregated in $\Alg$. Properties are
verified by translating them into a set of simple queries, which form a
sufficient condition for the target property to hold and which are sent
to the query solver. Importantly, \textsc{PM} attempts to infer certain
properties at a high level, in the absence of an index function, e.g.,
filtering/partitioning an injective array results in an injective array.
%filtering a monotonic array results in a monotonic array.
This enables scaling the analysis, e.g., across loops, which, unless trivial,
cannot produce useful index functions for their results.
{\em Finally}, the query solver (\textsc{QS}) uses the properties in $\Alg$ to
solve (in)equalities by extending Fourier-Motzkin elimination to work in
the presence of symbols such as array indexing and sums of array slices.

%Finally, the query solver (\textsc{QS}) uses the properties in $\Alg$ to
%solve (in)equalities in the $Poly$ representation.
%\textsc{QS} uses a strategy similar to Fourier-Motzkin elimination, but
%which is extended to work in the presence of symbols such as array indexing
%and  sums of array slices. %and exponential expressions.

%%%%%%%%%%%%%%%%%%%%%%%%%%%%%%%%%%%%%%%%%%
%%% Array Properties
%%%%%%%%%%%%%%%%%%%%%%%%%%%%%%%%%%%%%%%%%%
\subsection{Array Properties}
\label{subsec:array-props}

% (\backslash k \rightarrow e_k)

\begin{figure}
  \begin{minipage}[t]{0.99\linewidth}
  \begin{scriptsize}
  A segmented shape $(e^m,k,e_k)$ assumes that $0\leq k \leq e^m$, $e_k$ is monotonically increasing in $k$, and $e_0 = e_k\{k\mapsto 0\} = 0$.\vspace{-2ex}\\
   \[
    %\text{In the context of a segmented shape}~(e^m,k,e_k)~\text{it is assumed that} ~e_k~ \text{is monotonically increasing in $k$ and}~0\leq k \leq e^m\text{.}\newline %e_0=e_k\{k\mapsto 0\}=0\text{.}\\
    \textbf{Notation:} ~
      e_{k+1} =
        \begin{cases}
        e^n & \text{if} \hsp e^m = 0 \hsp \text{(only one segment)}\\
        e_k\{k\mapsto k+1\} & \text{if} \hsp e^m > 0 \hsp \text{(multiple segments)}
        \end{cases} \hsp \
        \RCD,\RCDimg=\begin{cases} \text{denotes an integral interval} & \\
                        \text{can be generalized, e.g, union of slices} &
          \end{cases}
   \]
  \end{scriptsize}
  \end{minipage}\vspace{-1ex}
  \begin{minipage}[t]{0.99\linewidth}
  \begin{scriptsize}
    \[
    \begin{array}{lcl}
    \textbf{Property} & \textbf{Known} & \textbf{Semantics}\\
    Monotonic^{\odot} \ X & X : [e^n]int &
      0 \leq i < j < e^n-1 \ \Rightarrow \ X[i] ~\odot~ X[j]\\
    \hsp \odot \in \{\leq, <, \geq, >\} & &\\
    Range \ X \ (e^{lb}, e^{ub}) & X : [e^n]int &
      0  \leq i < e^n \ \Rightarrow \ e^{lb} \leq X[i] \leq e^{ub} \medskip\\
    Equiv \ X \ e & & X \ \text{has an index function equivalent with the one derived from} \ e \medskip\\
    OrthogPreds \ & h = 1\ldots v &
                \forall (h_1 \not= h_2)\textbf{:}~0\leq k \leq e^m \ \wedge \ ~e_k \leq i < e_{k+1} \Rightarrow \neg (p_{h_1}(i) \wedge p_{h_2}(i))\\
                %(0\leq k \leq e^m \ \wedge \ ~e_k \leq i < e_{k+1} \ \Rightarrow \ p_1(i) \vee \ldots \vee p_v(i)) \ \ \wedge\\
    \hsp \textcolor{gray}{(e^m, k,e_k)} \ \overline{p_h} &  & \medskip\\
    Inj \ X \ \RCD & X : [e^n]int &
      (0 \leq j < i < e^n~\wedge~X[i] \in \RCD ~\wedge~ X[j] \in \RCD) ~ \Rightarrow ~ X[i] \not= X[j]\medskip\\
    Bij \ X \ \RCD & X: [e^n]int & Inj ~ X ~ \RCD \ \wedge \\
    \hsp (\textcolor{gray}{e^m, k, e_k,} \RCDimg_k) & & \RCDimg_k \overset{\text{set}}{=}  \{X[i] \ | \ e_k \leq i < e_{k+1} ~\wedge~ X[i] \in \RCD \} \medskip\\
% \mathlarger{\stackrel{\text{set}}{=}}
    FiltPart \ Y \ X  & X : [e^n]\tau &
                OrthogPreds \ (e^m,k,e_k) \ (p^p_1,\ldots,~p^p_v) \ \ \ \wedge \hfill\\
    \ (\textcolor{gray}{e^m, k, e_k,} p^f, \overline{p^p_h}) & Y:[e^y]\tau & Y \ \equiv \ \kw{map} \ ( \ \lambda k \rightarrow \ \kw{let} \ \sigma ~=~ \kw{filter} ~ p^f ~ [e_k, .., e_{k+1}-1] \hfill \\
    & h = 1\ldots v & \hsp\hsp\hsp\hsp\hsp\hsp\hsp \ \
                \kw{let} \ \sigma' \ = \ \kw{partition}_v \ (\overline{p^p_h}) \ \sigma \hfill \\
    &  & \hsp\hsp\hsp\hsp\hsp\hsp\hsp \ \
    \kw{in} \ \kw{map} \ (\lambda i\rightarrow X[i]) \ \sigma' \ ) \ [0, .., e^m-1] \ \ |> \ \kw{flatten}\medskip\\
    InvFiltPart \ X \ \RCDimg & X : [e^n]int &
         OrthogPreds \ (e^m,k,e_k) \ (p^p_1,\ldots,~p^p_v) \ \ \ \wedge \hfill\\
    \ (\textcolor{gray}{e^m, k, e_k,} p^f, \overline{p^p_{h'}}) & h'=1..v\text{-1} &
         Bij \ X \ \RCDimg \ ((0,k,0),~\RCDimg) \ \wedge \
         |\RCDimg| ~=~ \Sum{k}{0}{e^m}{\Sum{i}{e_k}{e_{k+1}-1}{p^f(i)}} \ \wedge
    \\
    & p_v = \neg (p_1 \vee&
         (~0\leq k_1 < k_2 \leq e^m \wedge
         e_{k_1} \leq j_1 < e_{k_1+1}\leq e_{k_2} \leq j_2 < e_{k_2+1} \wedge p^f(j_{1,2}) \ \Rightarrow \ Q) \ \wedge
    \\
    & \ \ldots \vee p_{v-1})&  (\forall h: ~ 0 \leq k \leq e^m \wedge e_k \leq j_1 < j_2 < e_{k+1} \wedge
         p^p_h(j_{1,2}) \wedge p^f{j_{1,2}} \ \Rightarrow \ Q) \ \wedge
    \\
    & h = 1\ldots v & (\forall(h_1<h_2): ~ 0 \leq k \leq e^m \wedge e_k \leq j_{1,2} < e_{k+1} \wedge
                p^p_{h_1}(j_1) \wedge p^p_{h_2}(j_2) \wedge p^f(j_{1,2})\ \Rightarrow \ Q)
    \\
    & & \textbf{where} \ Q \ \text{denotes the query} \ X[j_1] < X[j_2]
%    \bigskip\\
%    FiltPart \ Y \ X  & Y : [e^n]\tau &
%                InvFiltPart \ X \ [0,e^n) \ (e^m, k, e_k, p^f, \overline{p^p_h}) \hfill
%    \\
%    \ (e^m, k, e_k, p^f, \overline{p^p_h}) & X = [0\ldots n-1]&
    \end{array}
    \]
  \end{scriptsize}
  \end{minipage}\vspace{-2ex}
\caption{Array Properties. $X,Y,\sigma$ are array variables; $p$ denotes a predicate in lambda form of type $int \rightarrow bool$.}
\label{fig:array-props}
\end{figure}

%Notation: $v$ denotes an integral constant, $i,j,k$ denote fresh variable names; $X,Y,\sigma$ are variable names denoting arrays; $e$ denotes scalar (source) expressions that are translatable to $SoP$ representation; $p$ denotes a predicate in lambda form of type $int \rightarrow bool$; $f$ denotes an index function.

%\textcolor{gray}{

Figure~\ref{fig:array-props} presents the array properties supported by
our system. We ignore, for the moment, the gray text,
\textcolor{gray}{$e^m, k, e_k$}, which will be explained later.
Increasing and decreasing (strict) monotonicity is standard, as is the range
of an array element, except that the bounds are in polynomial representation.
$Equiv$ expresses equivalences, and it is mostly used for scalars,
e.g., the post-condition \lstinline{m == sum ( map i64.bool (map p xs) )} at
line~\ref{line:postcond-m} in~\cref{fig:running-egs}. $OrthogPreds$ requires
that the argument predicates are pairwise mutually exlcusive. The implementation
supports simple predicates by solving queries as directed by their index functions.
% the implementation supports simple predicates

% $Inj \ X \ \RCD$ says that filtering out the indices of $X$ that fall
% outside the restricted co-domain $\RCD$ results in an injective subarray.
$Inj \ X \ \RCD$ says that array $X$ is injective,
if it is restricted to the indices that map to values
within the \emph{restricted co-domain} $\RCD$ (specified as a range).
This is motivated by \lstinline{scatter dst is _} which ignores the values
in $is$ that are outside the bounds of \lstinline{dst}; if all
out-of-bounds values in \lstinline{is} are $\infty$, and
the remaining values are unique, then $Inj \ is \ [-\infty, \infty)$ holds
and \lstinline{scatter} is safe.
$Bij \ X \ \RCD \ \RCDimg$ similarly says that restricting $X$ to co-domain
$\RCD$ results in a subarray that is bijective in $\RCDimg$. It follows that
$\RCDimg \subseteq \RCD$ and the bijective subarray has no values in
$\RCD - \RCDimg$.

$FiltPart \ Y \ X \ p^f \ \overline{p^{p}}^{v-1}$ declares that the array $Y$ is
equivalent to filtering (the indices of) $X$ with %predicate
$p^f ~: ~int \rightarrow bool$ and then
performing a $v$-way partitioning with the pairwise mutually exclusive predicates
$\overline{p^p}$,
i.e., the elements that succeed under $p^p_1$ come before the ones that
succeed under $p^p_2$ and so on. By convention, if no filtering is performed,
then $p^f \_ = \True$ and an unknown filtering is represented by
$p^f \_ = \False$. $\overline{p^p}$ is similar, except
that an empty sequence also means unknown.
% The verification of this property is derived
%in a straightforward manner from the next property.

Verification of the previous property is enabled by
$InvFiltPart \ Z \ [0,e^n) \ p^f \ \overline{p^{p}}^{v-1}$,
which essentially declares that $Z$ is an array of
indices such that \lstinline{let Y = scatter Y0 Z X} (where $Y0 : [e^n]\tau$),
results in a filter-partitioning of $X$ with predicates $p^f$ and
$\overline{p^{p}}$, i.e., $FiltPart ~Y ~ X ~ p^f ~ \overline{p^{p}}^{v-1}$.
The semantics is that
(1) $Z$ restricted to co-domain $[0,e^n)$ is bijective in interval image $[0, e^n)$,
(2) the number of indices that succeed under $p^f$ equals $e^n$,
(3) for any $h=1\ldots v$, the indices that succeed under both $p^f$ and $p^p_h$ have
monotonically increasing values, and (4) for all pairs $(h_1 < h_2)$ the
values of the indices succeeding under $p^f$ and $p^p_{h_1}$ are smaller than
the ones that succeed under $p^f$ and $p^h_2$.
For example, in function \lstinline{partition2} of~\cref{fig:running-egs},
variable \lstinline{indices} at line~\ref{line:inds-of-scatter} has index function:\smallskip

$
\ixfn{\iot{i}{n}}{(cs[i] \Rightarrow -1 + \sum_{j=0}^i(cs[j])) ~\wedge~ (\neg cs[i] \Rightarrow i+\sum_{j=1+i}^{n-1}(cs[j]))} \hsp \text{where} ~cs[i] = p~ (x[i])
$\smallskip\\
which can be shown to satisfy
$InvFiltPart \ \text{indices} \ [0,n) \ (\backslash \_ \rightarrow \True) \ (\backslash i \rightarrow p~x[i])$,
i.e., the partitioning predicate is identified from the
guards. The next line performs the scatter and yields the result of
\lstinline{partition2} on which the filtering-partitioning post-condition holds.

The gray text \textcolor{gray}{$e^m, k, e_k$} in \cref{fig:array-props}
extends the properties $OrthogPreds$, $Bij$, $FiltPart$ and $InvFiltPart$ to
cover the segmented case, i.e., where $k$ is a bound variable taking values in
$0\ldots e^m$, and $e_k$ is a monotonically increasing sequence in $k$ that
denotes the union of segments. For example, $Bij \ X \ \RCD \ (e^m, k, e_k, \RCDimg_k)$
denotes that restricting $X$ to co-domain $\RCD$ results in a per-segment
bijective image $\RCDimg$, where $\RCDimg$ may depend on $k$, but $\RCD$ does not.
In the source language, the post/preconditions referring to segmented arrays
require to bind $k$ as a lambda argument, e.g, a bijective precondition
on argument $X$ is expressed as:
$Bij \ X \ (e_{\RCD}^{lb}, e_{\RCD}^{ub}) \ (e^m, \backslash k \rightarrow (e_k, e_{\RCDimg}^{lb}, e_{\RCDimg}^{ub}))$.

Possible generalizations include, for example, extending the interval co-domains
to a finite union of slices or {\sc lmad}s~\cite{LMAD,futhark-sc22mem} or allowing
the range of an element to depend on its index.
%, or (3) allowing a segmented definition of $\RCD$.
%---but the current cases do not require them.

%%%%%%%%%%%%%%%%%%%%%%%%%%%%%%%%%%%%%%%%%%%%%%%%%
%%% section on array properties
%%%%%%%%%%%%%%%%%%%%%%%%%%%%%%%%%%%%%%%%%%%%%%%%%

\section{Verifying Array Properties from their Index Function}
\label{sec:prove-props}

%Discussion is organized as follows:
\Cref{subsec:notation} introduces notation and the rationale of the design,
\cref{subsec:inj-bij-verif} presents the verification of injective, bijective
and filtering-partitioning properties (simpler properties such as ranges and
monotonicity are omitted for brevity), and \cref{subsec:new-prop-hl} presents
further automation for inferring properties at a high level, in
the~absence~of~an~index~function.

\subsection{Rationale of the Design and Notation}
\label{subsec:notation}

The rationale of the design is to define a small set of properties that
\begin{enumerate*}
\item are accessible to the non-expert user under a gentle learning curve,
\item are known to the compiler, and
\item expose a compositional algebra that allows the compiler to scale/automate
  the analysis as much as possible without the user's intervention.
\end{enumerate*}
However, the user's involvement is key to verification, not only in specifying
the properties of the code, but also in performing strategic
modifications/annotations that are rooted in the observation that it is much
easier to verify a property than to infer it: e.g., it is easier to verify that
array elements are within a given range than to infer the range.

As such, the user can guide the analysis by breaking a program into multiple
functions that are annotated with pre- and postconditions---including
equivalences on array sizes that enable unification of properties across
if-expressions.  This is facilitated by an architectural design that provides
facilities to inspect the relevant index functions and environment in order
to reason about whether a given property is actually provable or additional
properties need to be specified.

{\em The paper uses the following notation:}
\begin{itemize}
\item $\FV{o}$/$\BV{o}$ denote the free/bound variables of an object $o$.
      We write ``$o~\text{unifies with}~o'$'' if $o$ and $o'$ are
      syntactically identical up to the names of bound
      variables~\cite{sieg1993unification}.
      Bound variables appear in sums and segmented domains, for example,
      $x_1$ is bound in $\Sum{x_1}{e_1}{e_2}{s_1}$.
      %\nh{TODO Later on explain unification with contexts.}

\item We use $\Env; \Alg \vd f \leadsto f'$ and $\Env; \Alg \vd e \leadsto e'$
      to denote simplification of index functions and expressions.
      We assume that both are already in simplified form
      (see \cref{sec:query-solver}), and operations and substitutions such as
      $e_1-e_2$ and $e\{k\mapsto e_k\}$ also simplify the result.
      %include simplification of the result.
      % $\Sigma; \Alg \vd e \leadsto e'$ and of internal-language expressions and ;

\item The algebraic environment $\Alg$ is seen as a record whose fields
      are symbol tables, named after the corresponding properties, e.g.,
      $\Alg.Inj$ denotes injectivity and the new environment created by
      adding binding $x\mapsto \RCD$ to $\Alg.Inj$ is denoted by:
      $\Alg~\text{with}~Inj=\Alg.Inj ~\cup~\{x\mapsto \RCD\}$.

\item In other places we are less explicit:
      $\Alg \wedge (e^x = e^y) \wedge (e_1 \leq e_2 < e_3)$ denotes
      extending the equivalence symbol table $\Alg.Equiv$ with bindings
      derived from $e^x = e^y$ and the ranges table $\Alg.Range$
      with the bindings derived from inequalities $e_1 \leq e_2$,
      $e_2 < e_3$ and $e_1 < e_3$, as explained in \cref{sec:query-solver}.
      The last inequality $e_1 < e_3$ is not necessarily subsumed by the other two,
      e.g., assuming a positive \lstinline{shape} and $k$,
      $\sum_{k'=0}^{k-1} (\text{shape}[k']) \leq j < \sum_{k'=0}^{k} (\text{shape}[k'])$
      would result in the last inequality being simplified to
%      $\sum_{k'=0}^{k-1} (\text{shape}[k'] < \sum_{k'=0}^{k} (\text{shape}[k'])$,
%      which simplify to
      $\text{shape}[k] > 0$, which would improve the
      lower bound of symbol $\text{shape}[k]$ to be $1$ rather than $0$.
\item For convenience, we also define the shorthand notation below
      for extending $\Alg$ with the inequalities assumed by $ixfn$'s domains,
      where $e_{k+1} = e_k\{k\mapsto k+1\}$ and $e_{sz} = e_{k+1} - e_k$:
      \vspace{-1ex}
\[\hfill\hspace{6ex}
\begin{array}{lcl}
  \ranges{\Alg}{\forix~{\iot{i}{e_n}}} = \Alg \land 0 \leq i < e_n
  & \text{and} &
%  \ranges{\Alg}{\cat{k=0}{e_m}\seg{j}{e_{k}}} =
  \Alg^{+\cup_{k=0}^{e^m} \seg{j}{e_{k}}} =
  \Alg \land 0 \leq k \leq e_m \land e_k \leq j < e_{k+1} \land 0 \leq e_{sz}
\end{array}
\]\vspace{-3ex}

\item Queries are marked in \colorbox{black!6}{gray}, use the notation
      \colorbox{black!6}{$\Alg \vd c^1 \? c^2$} and succeed when it
      can be shown that $c^1$ implies $c^2$ in context $\Alg$.
      Queries proceed by
      converting $c^1$ to DNF (i.e.,  $c^1 = c_1^1 \vee \ldots \vee c_q^1$),
      and then asking the solver to prove $c^2$ in contexts
      $\Alg \wedge c^1_i$ where $i=1\ldots q$. The query succeeds if all
      $q$ sub-queries succeed.
\end{itemize}

%%%%%%%%%%%%%%%%%%%%%%%%%%%%
%%% BEG INJ/BIJ FIGURE
%%%%%%%%%%%%%%%%%%%%%%%%%%%%

\begin{figure}
  \sempart{Verifying bijectivity of a variable $x$ in restricted co-domain.}{\Env; \Alg \vd (x, \RCD, (k,\RCDimg)) \To{Bij} (\True,\Alg')}
    \begin{scriptsize}
    \begin{minipage}[t]{0.44\linewidth}
    \vspace{-3ex}
    \begin{align*}
    \inference{
      \Alg.Bij(x) = (\RCD_2, (\text{Sgm}_2,\RCDimg_2)) \\
      \text{Sgm}_2 = (e^m,k,e_k)\\
      \RCDimg_2 \text{ unifies with } \RCDimg_1\{k'\mapsto k\}\\
      \text{\colorbox{black!6}{$\Alg \vd 0\leq k \leq e^m \? \RCDimg_2 \subseteq \RCD_1 \subseteq \RCD_2$}}
    }{
      \Env; \Alg \vd  (x, \RCD_1, (k',\RCDimg_1)) \To{Bij} (\True,\Alg)
    }
    \tagsc{BijV1}
    \end{align*}
    \end{minipage}\begin{minipage}[t]{0.55\linewidth}
    \vspace{-1ex}
    \begin{align*}
    \inference{
      e^n = \len{x} \hsp\hsp\hsp\hsp \fresh{i}
      \\
      f = \ixfn{\iot{i}{e^n}}{(x[i] \in \RCD => x[i])} \gdelim (x[i] \notin \RCD => \infty)
      %\phantom{f = \ixfn{\iot{i}{e^n}}{}} \gdelim (x[i] \notin \RCD => \infty)
      \\
      \Env; \Alg \vd f \leadsto f'
      \hsp
      \Alg \vd ((k,\RCDimg),f') \To{Bij} (\True, (\text{Sgm}',\RCDimg'))
      \\
      \Alg' = \Alg ~\text{with}~Bij=\Alg.Bij \cup \{x \mapsto (\RCD, (\text{Sgm}',\RCDimg'))
    }{
      \Env; \Alg \vd (x, \RCD, (k,\RCDimg)) \To{Bij} (\True, \Alg')
    }
    \tagsc{BijV2}
    \end{align*}
    \end{minipage}
    \end{scriptsize}%\bigskip

  \sempart{Verifying bijectivity in image of an index function $f$.}{\Alg \vd  ((k,\RCDimg), f) \To{Bij} (\True, (\text{Sgm},\RCDimg))}
  \begin{scriptsize}
    \begin{align*}\vspace{-1ex}
    \inference{
      D=\union{k}{0}{e^m}{\text{for}~i \geq e_k}
        \hsp
        k' \not\in \mathcal{FV}(e^{lb}) \cup \mathcal{FV}(e^{ub})
        \hsp
        \hsp\hsp
        \text{\colorbox{black!6}{$\forall h \in 1,\dots,v ~: \
          \ranges{\Alg}{\forix~D} \vd c_h \? e^{lb} \leq e_h \leq e^{ub}$}}
      \vspace{-1ex}\\
      %\ranges{\Alg}{\forix~D}
      \Alg \vd f \To{Inj} \True
        \hsp\hsp
        e_{k+1} = e_k\{k\mapsto k+1\}
        \hsp\hsp
        \text{\colorbox{black!6}{$\Alg \vd \True \ \? \ e^{ub} + 1 - e^{lb} = \Sum{k}{0}{e^m}{\Sum{i}{e_k}{e_{k+1}-1}{c_1 \lor \ldots \lor c_v}}$}}
    }{
      \Alg \vd (
        (k', [e^{lb},e^{ub}]),
        f = \ixfn{D}{
                \ges{h}{1}{v}{c_h => e_h}
                \textcolor{gray}{[\gdelim (c^{\infty} => \infty)]}
              }
      ) \To{Bij} (\True, ((0, k', 0),~ [e^{lb},e^{ub}]))
    }
    \tagsc{BijF1}
    \end{align*}\vspace{-2ex}
    \begin{align*}
    \inference{
      i, \text{Sgm}, e_{k+1},\RCDimg = %\hfill
      %\\
      %  \hsp\hsp
        \text{$\begin{cases}
%        i,(e^m,k,e_k), ~e_{k+1}, ~[e_k, e_{k+1}) & \textbf{if} ~D=\union{k}{0}{e^m}{\text{for}~i \geq e_k}
%          \ \wedge \ \RCDimg'=[0,e'^n) \ \wedge \ e'^n ~\text{unifies with}~e^n
%        \\
%        & \hsp\hsp\hsp \textbf{where}~e_{k+1}=e_k\{k\mapsto k+1\},~e^n=e_k\{k\mapsto e^m+1\}\\
        i, (e^m,k,e_k), ~e_{k+1},~\RCDimg'\{k'\mapsto k\} & \textbf{if}~D=\union{k}{0}{e^m}{\text{for}~i \geq e_k},~\textbf{where}~e_{k+1}=e_k\{k\mapsto k+1\}\\
        i, (0,k',0), ~e^n, ~\RCDimg' & \textbf{if}~ D = \text{for}~i<e^n
        \end{cases}$}
      \\
      \Alg \vd f \To{Inj} \True
      \hsp\hsp\hsp\hsp
      \RCDimg = [e^{lb}, e^{ub}]
      \hsp\hsp\hsp\hsp
      \text{\colorbox{black!6}{$\forall h \in \{1,\dots,v\} :
        \ranges{\Alg}{\forix~D} \vd c_h \? e^{lb} \leq e_h \leq e^{ub}$}}
      \vspace{-1ex}\\
      \text{Sgm} = (e^m, k, e_k)
      \hsp e_{k+1} = e_k\{k\mapsto k+1\}
      \hsp\hsp\hsp\hsp
      \text{\colorbox{black!6}{$\ranges{\Alg}{\forix~D} \vd \True \? e^{ub} + 1 - e^{lb} = \Sum{i}{e_k}{e_{k+1}-1}{c_1 \lor \ldots \lor c_v}$}}
    }{
      \Alg \vd (
        (k', \RCDimg'),
        f = \ixfn{D}{
                \ges{h}{1}{v}{c_h => e_h}
                \textcolor{gray}{[\gdelim (c^{\infty} => \infty)]}
              }
      ) \To{Bij} (\True, (\text{Sgm}, \RCDimg))
    }
    \tagsc{BijF2}
    \end{align*}
  \end{scriptsize}
  \vspace{-1ex}

  \sempart{Verifying injectivity of a variable $x$ in a (restricted) co-domain.}{\Env; \Alg \vd (\RCD, x) \To{Inj} (true, \Alg')}
    \begin{scriptsize}
    \begin{minipage}[t]{0.33\linewidth}
    \begin{align*}
    \inference{
      % x \in \Env.Inj \\
      \RCD_2 = \Env.Inj(x)\\
      \text{\colorbox{black!6}{$\Alg \vd \True \? \RCD_1 \subseteq \RCD_2$}}
    }{
      \Env; \Alg \vd (\RCD_1, x) \To{Inj} (\True, \Alg)
    }
    \tagsc{InjV1}
    \end{align*}
    \end{minipage}\begin{minipage}[t]{0.69\linewidth}
    \vspace{-2ex}
    \begin{align*}
    \inference{
      \text{$\begin{array}{l}(i,e^n,\\ \ c,e^v)\end{array}$} \hspace{-2ex}=
        \text{$\begin{cases}
               (i, ~\text{length}~x, ~c^f \wedge x[i]\in \RCD, ~x[i]) &
                   \textbf{if} \ \Alg.FP(y)=(x,\lambda i.c^f,\_)\\
               (\fresh{i}, ~\text{length~y}, ~y[i]\in \RCD, ~y[i]) &
                   \textbf{otherwise}
               \end{cases}
             $}
      \\
      f = \ixfn{\iot{i}{e^n}}{(c => e^v) \gdelim (\neg c => \infty)}
      \hsp
      \Env; \Alg \vd f \leadsto f'
      \hsp
        \Alg \vd f' \To{Inj} \True
    }{
      \Env; \Alg \vd (\RCD, y) \To{Inj} (\True, \Alg~\text{with}~~Inj=\Alg.Inj \cup \{y \mapsto \RCD\})
    }
    \tagsc{InjV2}
    \end{align*}
%
%    \begin{align*}
%    \inference{
%      e^n = \len{x}
%        \hsp\hsp\hsp\hsp
%        \fresh{i}
%      \\
%      f = \ixfn{\iot{i}{e^n}}{(x[i] \in \RCD => x[i]) \gdelim
%                              (x[i] \not\in \RCD => \infty)}
%      \\
%      \Env; \Alg \vd f \leadsto f'
%        \hsp\hsp\hsp\hsp %\Env;
%        \Alg \vd f' \To{Inj} \True
%    }{
%      \Env; \Alg \vd (\RCD, x) \To{Inj} (\True, \Alg~\text{with}~~Inj=\Alg.Inj \cup \{x \mapsto \RCD\})
%    }
%    \tagsc{InjV2}
%    \end{align*}
    \end{minipage}
    \end{scriptsize}

    % \begin{scriptsize}
    % \nh{Cosmin: I need this rule for the equational proof. Can it be incorporated somehow?}
    % %
    % \begin{align*}
    % \inference{
    %   \Alg.FP(y) = (x, \lambda i'.~p^f, \_)
    %   \hsp
    %   \Env(x) = \ixfn{\iot{i}{e^n}}{\ges{h}{1}{v}{c_h => e_h}}
    %   \\
    %   \fresh{j}
    %   \hsp
    %   \ranges{\Alg}{\iot{i}{e^n},}\ranges{}{\iot{j}{e^n}} \vd_{i,j} {\ges{h}{1}{v}{c_h \land p^f\{i'\mapsto i\} => e_h}}
    %     \To{Inj=}~ \True
    % }{
    %   \Env; \Alg \vd (\RCD, y) \To{Inj} (\True, \Alg~\text{with}~~Inj=\Alg.Inj \cup \{y \mapsto \RCD\})
    % }
    % \tagsc{InjV3}
    % \end{align*}
    % \end{scriptsize}

  \sempart{Verifying injectivity of an index function $f$.}{\Alg \vd f \To{Inj} \True}
  \begin{scriptsize}
    \begin{minipage}[t]{0.37\linewidth}
    \begin{align*}
    \inference{
       f = \text{for}~j<e^n.~\overline{ge}~\textcolor{gray}{[\gdelim (\_ => \infty)]}
        %\bigwedge~(\_\Rightarrow \infty)
       \\
       \text{fresh}~j'
       \\
       \Alg' = \Alg \wedge 0 \leq j < e^n \wedge 0 \leq j' < e^n%\hfill
       \\
       \textbf{(} \hfill \Alg' \vd_{\not=,j,j'} \ \overline{ge} \ \To{Mon}~ (\kw{true}, \_)
       \\
       ~\textbf{OR} \hsp \Alg' \vd_{j,j'} \ \overline{ge} \ \To{Inj=}~ \kw{true} \hfill \textbf{)}
    }{
      \Alg \vd f \To{Inj} \True
    }
    \tagsc{InjF1}
    \end{align*}
    \end{minipage}\begin{minipage}[t]{0.64\linewidth}
    \vspace{-2ex}
    \begin{align*}
    \inference{
      fresh~k'~j'
        \hsp e_{k+1} = e_k\{k\mapsto k+1\}
        \hsp e'_{k} = e_k\{k\mapsto k'\}
      \\
      \Alg' = \Alg \wedge (0 \leq k \leq e^m) \wedge (e_k \leq j,j' < e_{k+1})
      \\
      \textbf{(} \ \Alg' \vd_{\not=,j,j'} \ \overline{ge} \ \To{Mon} (\True,\_) \hsp
        \textbf{OR} \hsp \Alg' \vd_{j,j'} \ \overline{ge} \ \To{Inj=} \True \ \textbf{)}
      \\
      \hsp e'_{k+1} = e_k\{k\mapsto k'+1\} \hsp\hsp
        \overline{ge}' = \overline{ge}\{j\mapsto j', k\mapsto k'\}
      \\
      \Alg'' = \Alg \wedge (0\leq k < k' \leq e^m) \wedge (e_k \leq j < e_{k+1}) \wedge (e'_{k} \leq j' < e'_{k+1})
      \\
      \Alg'' \vd_{\not=} (\overline{ge}, \overline{ge}') \To{Cmp} \True
    }{
       \Alg \vd \union{k}{0}{e^m}{\text{for}~ j \geq e_k~. \ \overline{ge}~\textcolor{gray}{[\gdelim (\_ => \infty)]}} \To{Inj} \kw{true}
    }
    \tagsc{InjF2}
    \end{align*}
    \end{minipage}
  \end{scriptsize}\vspace{-3ex}
  \caption{Verifying injectivity and bijectivity of variables and index functions denoting arrays.}
  \label{fig:ixfn-bij}
\end{figure}

%%%%%%%%%%%%%%%%%%%%%%%%%%%%
%%% END INJ/BIJ FIGURE
%%%%%%%%%%%%%%%%%%%%%%%%%%%%

%%%%%%%%%%%%%%%%%%%%%%%%%%%%%%%%%%%%%%
%%% Injectivity and Bijectivity
%%%%%%%%%%%%%%%%%%%%%%%%%%%%%%%%%%%%%%

\subsection{Verifying Injectivity and Bijectivity of Index Functions}
\label{subsec:inj-bij-verif}

Figure~\ref{fig:ixfn-bij} presents the inference rules that verify injective and
bijective properties.
% and \cref{fig:ixfn-mon-inj-helpers} shows helper function.
%
Rules \textsc{BijV1} and \textsc{BijV2} take the form
$\Env; \Alg \vd (x, \RCD, (k,\RCDimg)) \To{Bij} (\textbf{bool},\Alg')$
that answers whether the array denoted by variable $x$, when viewed
as an index function restricted
to the pre-image of $\RCD$, has bijective image $\RCDimg$; $k$ enables the
treatment of (jagged) segmented arrays, i.e., when $\RCDimg$ depends on $k$
then each segment (of index $k$) of the array has bijective image $\RCDimg$.
The result is a boolean ($\True$ means success) and a new symbol table,
that possibly extends $\Alg.Bij$ with a newly verified binding.

Rule \textsc{BijV1} requires that $x$ already has a binding, denoted
$(\RCD_2, (\text{Sgm}_2, \RCDimg2))$, where $\text{Sgm}_2$ denotes a
segmented shape $(e^m, k, e_k)$, as introduced in~\cref{fig:array-props} and
\cref{subsec:array-props}. The rule succeeds if (1) the queried
image $\RCDimg_1$ is equivalent to the one of the binding $\RCDimg_2$,
and (2) it can be proven that $\RCDimg_2 \subseteq \RCD_1 \subseteq \RCD_2$.\footnote{
Verification of $[e^{lb}_1, ~e^{ub}_1] \subseteq [e^{lb}_2, ~e^{ub}_2]$
is achieved by trying to prove
$(e^{lb}_1 > e^{ub}_1) ~\vee~ (e^{lb}_1 \geq e^{lb}_2 \ \wedge \ e^{ub}_1 \leq e^{ub}_2$).
%$\RCDimg_2 \subseteq \RCD_1 \subseteq \RCD_2$ is proven by splitting
%it into a set of basic inequality queries that are sent to the solver
%(not shown).
}
Since by construction $\RCDimg_2\subseteq \RCD_2$, it follows that $x$
can have no points in $\RCD_2 - \RCDimg_2$ therefore the restriction to
co-domain $\RCD_1 \subseteq \RCD_2$ also has bijective image $\RCDimg_1$.

Rule \textsc{BijV2} covers the case when $x$ does not have a %suitable
binding in $\Alg.Bij$. It creates an index function that maps all
points inside the queried $\RCD$ to itself (\lstinline{x[i]})
and the other points to a~special~$\infty$~symbol.
Then it simplifies it (see \cref{sec:infer-ixfn}) and
tries to prove bijectivity of the resulted~index~function.
%If successful it updates $\Alg.Bij$.

This is achieved by a judgment of form
$\Alg \vd  ((k,\RCDimg), f) \To{Bij} (\True, (\text{Sgm},\RCDimg))$
that similarly returns $\True$ for success, together with the (segmented)
bijective image; $\RCD$ is missing because all the points outside it
have been mapped to $\infty$.

Rule \textsc{BijF1} covers the case when the index function $f$ is segmented,
but $\RCDimg$, denoted by interval $[e^{lb}, e^{ub}]$, is not,
i.e., the image does not depend on $k$. Bijectivity is verified by checking
that (1) all guarded expressions that do not correspond to $\infty$ yield
values within $\RCDimg$, (2) $f$ is injective outside $\infty$ points,
%($\Alg \vd f \To{inj} \True$),
and (3) the cardinal of $\RCDimg$ is equal with the number of indices that
are not mapped to $\infty$.

Rule \textsc{BijF2} covers two other cases\footnote{
A special case (not shown) refers to when $f$ is segmented, $\RCDimg$
is equal to the whole (unsegmented) domain of $f$, and each segment
$[e_k,e_{k+1})$
contains values that are a permutation of $[e_k,e_{k+1})$.
The result is the segmented bijection $[e_k,e_{k+1})$.
}: %in a similar way:
one that refers to when both $f$ and $\RCDimg$ are segmented,
in which case bijectivity is verified for each segment of $f$,
%(by checking injectivity and that $\RCDimg$ has the same number of points
%as the ones of the segment that ae not $\infty$)
and another one, in which $f$ has the linear domain
and the property is considered non segmented.
Verification is similar to \textsc{BijF1}.
%The verification of bijectivity is similar to \textsc{BijF1}.

Rules \textsc{InjV1} and \textsc{InjV2} verify injectivity in
a restricted co-domain $\RCD$ of a variable $x$ and are
similar in the form of judgments and treatment with rules
\textsc{BijV1} and \textsc{BijV2}.
What differs is that \textsc{InjV2} also checks whether the
target array $y$ is a filtering partitioning of some array $x$,
in which case it reasons in terms of $x$ while taking into
consideration the filtering predicate.

%%%%%%%%%%%%%%%%%%%%%%%%%%%%
%%% BEG HELPER  FIGURE
%%%%%%%%%%%%%%%%%%%%%%%%%%%%

\begin{figure}
  \sempart{Guarded-Exps Helpers.}{\Alg \vd_{\odot,i,j} \ \overline{ge} \ \To{Mon} (\True, \sigma) \hsp \Alg \vd_{i,j} \ \overline{ge} \ \To{Inj=} \True \hsp \Alg \vd_{\odot} (\overline{ge}^1,~\overline{ge}^2) \To{Cmp} \True}
  \begin{scriptsize}\vspace{-2ex}
    \begin{align*}
    \inference{
      %fresh \ j \hsp\hsp %\Env' = \Env, ~ e^{lb} \leq j < i < e^{ub} \hsp\hsp
      \forall h \in 1\ldots v~\text{it holds:} \ \textbf{(} \hsp
          c_h' = c_h\{i\mapsto j\} \hsp
          e_h' = e_h\{i\mapsto j\} \hsp
          \text{\colorbox{black!6}{$\Alg \wedge i < j \vd \ c_h \wedge c'_h \ \? e_h ~\odot~e'_h$}}
          \hsp \textbf{)}\\
%            \text{for all} ~h \in 1\ldots v~\text{it holds that:} \hsp
%        \Env, e^{lb} \leq j < i < e^{ub} \vd^{Q} (c_h \wedge c'_h, \ \ e'_h < e_h \ \vee \ e'_h > e_h) \rightarrow \kw{true}\\
      \text{There exists a sorting permutation}~\sigma~\text{of}~\{1,\dots,v\}
        \text{ such that for all }
        h \in \{1,\dots,v\}~\text{with}~\sigma(h) < v:
      \\
      \hsp\hsp \text{\colorbox{black!6}{$\Alg \vd (j < i \ \vee \ i < j) \wedge c'_{\sigma(h)} \wedge c_{\sigma(h)+1} \ \? \ e'_{\sigma(h)} < e_{\sigma(h)+1}$}}
      %\hsp\hsp \Env' \vd^{mon}_{ges}
    }{
       \Alg \vd_{\odot,i,j} \ \fvseqCustom{h=1}{v}{c_h \Rightarrow e_h} \ \To{Mon} \ (\True,\sigma)
    }
    \tagsc{MonGe}
    \end{align*}\vspace{-6ex}\\

    \begin{minipage}[t]{0.45\linewidth}
    \begin{align*}
    \inference{
      \sigma = \{i\mapsto j\}
      \ \
      V = \{1,\dots,v\}
      \ \
      \forall h,l \in V \times V:
      \\
      % \Alg' \vd c_h \wedge c_l\{i\mapsto j\} \wedge e_h = e_l\{i\mapsto j\} \? i=j
      \text{\colorbox{black!6}{$\Alg \vd \ (e_h = \sigma(e_l)) \ \wedge \ c_h \ \wedge \ \sigma(c_l)  \ \? \ i=j$}}
    }{
      \Alg \vd_{i,j} \
        \ges{h}{1}{v}{c_h => e_h} \
        \To{Inj=} \ \True
    }
    \tagsc{InjGe}
    \end{align*}
%    \nh{add equitional solver query to InjGe explicitly?}
    \end{minipage}\begin{minipage}[t]{0.56\linewidth}
    \begin{align*}
    \inference{
%      g_1 = \ges{h}{1}{v}{c^1_h => e^1_h}
%      \hsp
%      g_2 = \ges{l}{1}{w}{c^2_l => e^2_l}
      \forall h, l \in \{1,\dots,v\} \times \{1,\dots,w\}:~
      \text{\colorbox{black!6}{$\Alg \vd c^1_h \wedge c^2_l \? e^1_h \odot e^2_l$}}
    }{
       \Alg \vd_\odot (\ges{h}{1}{v}{c^1_h => e^1_h}, \ \ges{l}{1}{w}{c^2_l => e^2_l}) \To{Cmp} \True
    }
    \tagsc{CmpGe}
    \end{align*}
    \end{minipage}
    \end{scriptsize}
    \vspace{-2ex}

  \sempart{Filtering-partitioning property of a variable.}{\Env; \Alg \vd x \To{FP} (\True, \Alg')}\vspace{-1ex}
    \begin{scriptsize}
    \begin{minipage}[t]{0.29\linewidth}
    \vspace{2ex}
    \begin{align*}
    \inference{
      \Alg.FP(y) = (x, \text{fp\_prop})
    }{
      \Env; \Alg \vd y \To{FP} (\True, \Alg)
    }
    \tagsc{FPV1}
    \end{align*}
    \end{minipage}\begin{minipage}[t]{0.57\linewidth}
    \begin{align*}
    \inference{
      \Env(y) = \text{for} ~i < e^y.~true => x[is^{-1}[i]]
        \hsp\hsp
        \RCDimg = [0,e^y)
      \\
      \Env; \Alg \vd  (~is,~\RCDimg~) \To{IFP} (\True, \Alg')
        \hsp\hsp
        \Alg'.IFP(is) = (\_, ~\text{prop})
    }{
      \Env; \Alg \vd y \To{FP} (\True,~ \Alg'~\text{with}~ FP = \Alg'.FP \vee \{y\mapsto (x, \text{prop})\})
    }
    \tagsc{FPV2}
    \end{align*}
    \end{minipage}
  \end{scriptsize}\vspace{-2ex}
\caption{Guarded Expressions Helpers \& Translating Filter-Partitioning Property to Inverse Filtering Partitioning}
\label{fig:ixfn-mon-inj-helpers}
\end{figure}

The rules \textsc{InjF1} and \textsc{InjF2}, of form
$\Alg \vd f \To{Inj} \textbf{bool}$, verify the injectivity of
the non-$\infty$ elements of index function $f$.
Rule \textsc{InjF1} treats index functions having linear domains
%$\text{for j<e^n$,
and uses the helper inference rules $\To{Inj=}$ and $\To{Mon}$
defined in \cref{fig:ixfn-mon-inj-helpers}.
Injectivity is verified in either one of two ways:\\
\noindent The first approach uses $\To{Inj=}$ to prove that
$x[i_1] = x[i_2]$ implies $i_1=i_2$ for the values of
the guarded expressions other than the one leading to $\infty$,
i.e., $\_ \Rightarrow \infty$. This is achieved by the
query-solver technique presented in \cref{subsec:eq-with-inj}.
The second approach uses $\To{Mon}$ to prove a
sufficient condition based on piecewise monotonicity, namely:
%\begin{itemize}
%\item[1.]
\textbf{\em If} for any guarded expression $c_h \Rightarrow e_h$
the $e_h$ values are distinct---which is
denoted by $\not=$ in $\vd_{\not=,j,j'}$ and typically
comes down to strict monotonicity---\textbf{\em and}
%\item[2.]
there exists a sorting permutation $\sigma$ that reorganizes the
guarded expressions such that their values always increase across them
\textbf{\em then} the values of those guarded expressions are
injective.
%\end{itemize}

Rule \textsc{InjF2} is concerned with index functions that have
segmented domains. It applies a similar reasoning to \textsc{InjF1}
within a segment, and in addition, it uses the
$\Alg \vd_{\not=} (\overline{ge},\overline{ge'})$ rule to prove that
%values are either strictly increasing or strictly decreasing across
%segments.
values belonging to different segments are different;
$e_1 \not= e_2$ is solved by checking $e_1 < e_2$~or~$e_1 > e_2$.
%i.e., for any $k_1>k2$ all values of segment $k_1$ are
%less than those of segment $k_2$ or the reverse.

Rules \textsc{FPV1} and \textsc{FPV2} in~\cref{fig:ixfn-mon-inj-helpers}
sketch the treatment of filtering-partitioning properties.
Since such properties can only result in an index function
of form $\text{for} ~i < e^y.~true => x[is^{-1}[i]]$, rule \textsc{FPV2}
pattern matches said form and tries to infer the inverse filter-partitioning
property ($IFP$) on $is$ in bijective image $[0,e^y)$. % ($\To{IFP}$)
$IFP$'s semantics was discussed in \cref{subsec:array-props}. 
Its implementation (not shown) is similar
to bijectivity rule \textsc{BijV2}, except that:
(1) index function $f$ is filtering out the points outside image $[0,e^y)$;
(2) the calls to rule \textsc{InjGe} ($Inj=$) are
    %$\Alg'\vd_{j,j'} \overline{ge} \To{Inj=} \True$ are
    eliminated from \textsc{InjF1} and \textsc{InjF2};
(3) the calls to rules \textsc{MonGe} ($Mon$) and \textsc{CmpGe} ($Cmp$)
    are instantiated with $<$ instead of $\not=$, to check strictly
    {\em increasing} monotonicity.
The predicates are obtained from the guards of the simplified $f$:
the filtering one by negating the guard of the $\infty$ value,
and the partitioning ones from the guards producing
legal indices, ordered by the $\sigma$ permutation
computed by rule \textsc{MonGe}. %$\To{Mon}$

\subsection{Inferring New Properties at a High Level}
\label{subsec:new-prop-hl}

\begin{figure}
%
%  \sempart{Inferring Index Functions from let (source) expressions.}{\Env; \Alg \vd e \To{e^{*}2f} (f,\Env',\Alg')}
%    \begin{scriptsize}
%    %
%    \begin{align*}\vspace{-1ex}
%    \inference{
%      \Env;\Alg \vd e_1 \To{e^{*}2f} (f_1, \Env_1, \Alg_1)
%        \hsp
%        \Env_2 = \Env \cup \{x \mapsto f_1\}
%        \hsp
%        \Env_2;\Alg_1 \vd (x,e_1) \To{\Alg U} \Alg_2
%        \hsp
%        \Env_2;\Alg_2 \vd \To{e^{*}2f} (f, \Env_3, \Alg_3)
%    }{
%      \Env;\Alg \vd \kw{let}~x~=e_1~\kw{in}~e_2 \To{e^{*}2f} (f, \Env_3, \Alg_3)
%    }
%    \tagsc{InfLet}
%    \end{align*}\vspace{-2ex}
%    \end{scriptsize}
%
  \sempart{A Few Samples of Inferring Properties at a High Level.}{\Env; \Alg \vd (y, e^0) \To{\Alg U} \Alg'}
    \begin{scriptsize}
    \begin{minipage}[t]{0.57\linewidth}
    \vspace{-1ex}
    \begin{align*}
    \inference{
        \Alg.FP(y) = (x, \ \lambda i. e^f, \ \ldots)
      \ \
      e^f = \True
      \ \
      \Alg.Bij(x) = \text{RcdSgIm} \hfill
    }{
      \Env; \Alg \vd  (y, e) \ \To{\Alg U} \
        \Alg~\text{with}~Bij=\Alg.Bij \cup \{y\mapsto \text{RcdSgIm}\}
    }
    \tagsc{$\Alg$UeBij}
    \end{align*}
    \end{minipage}
%    \begin{minipage}[t]{0.57\linewidth}
%    \vspace{-1ex}
%    \begin{align*}
%    \inference{
%        \Alg.FP(y) = (x, \ \lambda i. e^f, \ \ldots)
%      \\
%      e^n = \text{length}~x
%        \hsp\hsp\hsp
%        \Alg.Bij(x) = (\RCD, ((\_,k,\_), \RCDimg))
%      \\
%      f = \text{for}~i<e^n.~(e^f \wedge x[i]\in \RCD \Rightarrow x[i]) \bigwedge
%      \\ \hspace{13ex}
%         (\neg e^f \vee x[i]\not\in \RCD \Rightarrow \infty)
%      \\
%      \Env;\Alg \vd f \leadsto f'
%        \hfill
%        \Alg \vd ((k,\RCDimg), f') \To{Bij} (\True, \text{SgIm})
%    }{
%      \Env; \Alg \vd  (y, e) \ \To{\Alg U} \
%        \Alg~\text{with}~Bij=\Alg.Big \cup \{y\mapsto (\RCD, \text{SgIm})\}
%    }
%    \tagsc{$\Alg$UeBij}
%    \end{align*}
%    \end{minipage}
    \begin{minipage}[t]{0.42\linewidth}
    \vspace{-1ex}
    \begin{align*}
    \inference{
      \overline{y}^v = \text{res}(e^1)
        \hsp\hsp
        \overline{z}^v = \text{res}(e^2)\\
      \Env;\Alg \vd (x_1, x^c, y_1, z_1) \To{IfFP} \Alg_1
        \hsp \ldots
      \\
      \Env;\Alg_{v-1} \vd (x_v, x^c, y_v, z_v) \To{IfFP} \Alg_v
      \\
      \Env;\Alg_{v} \vd (\overline{x}^v, x^c, \overline{y}^v, \overline{z}^v) \To{IfBij} \Alg'
        \hsp (\ldots)
    }{
      \Env; \Alg \vd  (\overline{x}^v, \kw{if}~x^c~\kw{then}~e^1~\kw{else}~e^2) \ \To{\Alg U} \ \Alg'
    }
    \tagsc{$\Alg$Uif}
    \end{align*}
    \end{minipage}\medskip

    \begin{minipage}[t]{0.52\linewidth}
    \vspace{-10ex}
    \begin{align*}
    \inference{
%      |\overline{y}| = |\overline{x^1}| = |\overline{x^2}| = v
%      \\
      \textbf{FOR ANY} \ \ h=1\ldots v \ \ \textbf{SUCH THAT:}\hfill
      \\
      \hsp\hsp
        (R^1, \text{SgIm}^1) = \Alg.Bij(x^1_h)\overline{\{x^1_l \mapsto y_l\}}^{l=1\ldots v} \hfill
      \\
      \hsp\hsp
        (R^2, \text{SgIm}^2) = \Alg.Bij(x^2_h)\overline{\{x^2_l \mapsto y_l\}}^{l=1\ldots v} \hfill
      \\
      \hsp
      \text{$\begin{array}{c}
      \text{SgIm}^1 ~\text{unifies}\\
      \text{with} ~ \text{SgIm}^2
      \end{array}$}
      \hsp
      \text{$R = \begin{cases}
        R^1 & \textbf{if} \ \text{\colorbox{black!6}{$\Alg \vd \True \? R^1 \subseteq R^2$}}\vspace{-1ex}\\
        R^2 & \textbf{if} \ \text{\colorbox{black!6}{$\Alg \vd \True \? R^2 \subseteq R^1$}}
        \end{cases}$}
      \\
      \textbf{UPDATE} \ \ M \leftarrow M \cup \{y_h \mapsto (R,SgIm^1)\}\hfill
    }{
      \Env; \Alg \vd  (\overline{y}^v, \_, \overline{x^1}, \overline{x^2}) \ \To{IfBij} \
        \Alg~\text{with}~Bij=\Alg.Big \cup M
    }
    \tagsc{IfBij}
    \end{align*}
    \end{minipage}\begin{minipage}[t]{0.47\linewidth}
    \begin{align*}
    \inference{
      \Alg.FP(x^1) = (z^1, p_0^1, \overline{p^1_h}^{h=1..v_1}) \hfill
      \\
      \Alg.FP(x^2) = (z^2, p_0^2, \overline{p^2_h}^{h=1..v_2})
      \hfill z^1 = z^2 \hsp  v_1 = v_2
      \\
      \textbf{for all}~h=0\ldots v_1:
        \Env;\Alg \vd (x^c, p_h^1, p_h^2) \To{predU} p_h
      \\
      \Alg' = \Alg~\text{with}~FP=\Alg.FP\cup \{y \mapsto (z_1, \overline{p_h}^{h=0..v_1})\}
    }{
      \Env; \Alg \vd  (y, x^c, x^1, x^2) \ \ \To{IfFP} \ \ \Alg'
    }
    \tagsc{IfFP}
    \end{align*}
    \end{minipage}\medskip

    \begin{minipage}[t]{0.47\linewidth}
    \vspace{-3ex}
    \begin{align*}
    \inference{
      p^1 = \lambda i.~ e^1
      \hsp\hsp
      p^2 = \lambda i.~ e^2
      \hsp\hsp
      e^{1,2} \not= \textbf{false}
      \\
      \Env;\Alg \vd (x^c \wedge e^1) \ \vee \ (\neg x^c \wedge e^2\{i'\mapsto i\}) \leadsto e
    }{
      \Env; \Alg \vd  (x^c, p^1, p^2) \ \To{predU} \ \lambda i.~ e
    }
    \tagsc{PredUS}
    \end{align*}
    \end{minipage}\begin{minipage}[t]{0.52\linewidth}
    \begin{align*}
    \inference{
      p^1 = \lambda i.~ \textbf{false}
      \hsp\hsp\hsp \textbf{OR} \hsp\hsp\hsp
      p^2 = \lambda i.~ \textbf{false}
    }{
      \Env; \Alg \vd  (x^c, p^1, p^2) \ \To{predU} \ \lambda i.~ \textbf{false}
    }
    \tagsc{PredUF}
    \end{align*}
    \end{minipage}
    \end{scriptsize}\vspace{-2ex}
  \caption{Inferring and Verifying Properties at a High Level.}
  \label{fig:infer-hl}
\end{figure}

Figure~\ref{fig:infer-hl} demonstrates several illustrative
rules for inferring %bijectivity and filtering-partitioning
properties at a high level.
Judgments take the form
$\Env;\Alg \vd (y, e^0) \To{\Alg U} \Alg'$,
in which the argument indicates a source-language binding
$\kw{let}~ y = e^0$ and the result is a potentially extended
symbol table.

Rule \textsc{$\Alg$UeBij} states that if $y$ is a partitioning
of $x$ and $x$ is bijective in a restricted co-domain, then the
bijectivity property is transferred to $y$. Other (not shown)
properties are similarly derived, e.g., filtering preserves
monotonicity and filtering-partitioning preserves injectivity,
and can be used to refine ranges
(by min/maxing the existent upper/lower bounds of $y$ with those of $x$).

%treats the case of a variable
%$y$, which is known to be a filtering partitioning of a
%variable $x$, which is known to have bijective image $\RCDimg$.
%The rule checks whether the bijective image still holds for
%$y$ by essentially checking that $p^f(i)$ implies $x[i] \not\in \RCD$,
%where $p^f$ denotes the filtering predicate---i.e., this guarantees
%that all filtered elements do not belong to the image.
%If this is not the case and bijectivity holds on $y$ in an
%expressible image, then user intervention is necessary to
%specify it.
%
%Other properties are derived in similar, but straightforward ways,
%such as: (1) filtering partitioning property preserves injectivity,
%and can be used to refine the range of $y$ with the range of $x$
%by min/maxing the existent upper/lower bounds of $y$,
%(2) filtering preserves monotonicity, and
%(3) partitioning preserves bijectivity.

Rule \textsc{$\Alg$Uif} aims to unify properties across if
expressions, which is demonstrated for bijectivity
(\textsc{IfBij}) and filtering-partitioning properties (\textsc{IfFP}).
Rule \textsc{IfBij} denotes by $\overline{y}$ the variable bound to the
if expression and by $\overline{x^1}$ and $\overline{x^2}$
the variable results of the then and else expressions, respectively,
all of them necessarily having the same (tuple) cardinality $v$.
If for any $h \in 1 \ldots v$ it happens that both $x^1_h$ and
$x^2_h$ are bijective, with equivalent images
and restricted co-domains
that are in an inclusion relation, then the bijective property
is transmitted to the corresponding if result by choosing the smaller
restricted co-domain.
The rule uses an improvement that substitutes the results of the
then/else branches for the ones of the if result in the bijective
property before unifying them. This enables the human in the loop,
i.e., even if the images are not equal, the user can ``force'' unification
by wrapping the then/else expressions into functions whose post-conditions
use dependent typing across results---e.g., if both function have
post-condition $\lambda n ~X.~Bij \ X \ (0, 8*n) \ (\_, (n,2*n))$
then bijectivity will unify even if the $n$ result might differ across
branches.

Rule \textsc{IfFP} uses a notation similar to \textsc{IfBij}, except that,
for simplicity, it does not use the dependent-typing refinement, i.e.,
$y,x_1,x_2$ denote single variables. In addition, it denotes with $x^c$
the if condition. The rules states that if both $x^1$ and $x^2$ are filtering
partitioning of the same array (variable $z$) with a matching number of
partitions $v$, then they can be accurately unified across the if expression.
The resulting predicates are computed by \textsc{PredUS} as
$p~ i = (x^c \wedge p^1(i)) \vee (\neg x^c \wedge p^2(i))$,
where $p^1$ and $p^2$ corresponds to the predicates of $x^1$ and $x^2$,
unless one of them denotes unknown ($\lambda i.\textbf{false}$), in which
case they unify as unknown by rule \textsc{PredUF}. If the number of
partitions do not match, then they unify as a fully-unknown partitioning,
denoted by an empty sequence of predicates~(not~shown).

%\newpage

%%%%%%%%%%%%%%%%%%%%%%%%%%%%%%%%%%%%%%%%%%%%%
%%% Analysis for Inferring Index Functions
%%%%%%%%%%%%%%%%%%%%%%%%%%%%%%%%%%%%%%%%%%%%%

\section{Inferring Index Functions}
\label{sec:infer-ixfn}

\begin{figure}
  \input{deps/convert.tex}\vspace{-2ex}
  \caption{Converting the source language to index functions.}
  \label{fig:convert}
\end{figure}

Our source programs are compositions of array combinators, encouraging a
point-free programming style. However, pointful reasoning naturally emerges
as a proof discipline for proving properties. For instance, to prove
$Inj~xs~(-\infty,\infty)$ it is sufficient to verify the query
% \[
%  \bigwedge~\left\{0 \leq i < \len{xs},~0 \leq j < \len{xs},~ xs[i] = xs[j]\right\} \? i = j.
% \]
\[
 \Alg \vd \lr{0 \leq i < \len{xs} ~\land~ 0 \leq j < \len{xs} ~\land~ xs[i] = xs[j]} \? i = j.
\]
To bridge the gap between program and query, we do away with the abstraction
provided by array combinators by transforming source programs to index
functions, enabling precise reasoning about array elements.
\Cref{fig:convert} provides an initial set of rules for inferring index
functions (excluding $\kw{scan}$ and $\kw{scatter}$). For simplicity,
all variables are treated as flat arrays (scalars are single-element arrays),
the rules target a subset of the language without tuples (SoA),
and we sometimes write\ $~\scalar~$~\ for $\iot{i}{1}$.
Function declarations are translated similarly to
let-bindings, with these additions:
\begin{enumerate*}
  \item index functions are created for the formal arguments and bound in $\Env$;
  \item preconditions on arguments are added to $\Alg$; and
  \item the postcondition is verified on the resulting index functions.
\end{enumerate*}

We demonstrate the rules on our running example \lstinline{partition2}
from \cref{fig:running-egs}.
Recall that the example takes a predicate $p$ as argument, which we treat as an
uninterpreted function by inserting an indexing symbol. % here (and in the implementation).
The first source expression \lstinline{let cs = map (\x -> p x) xs} is transformed by:
\begin{small}
\begin{align*}
\inference{
  \inference{
    \text{$p$ has type $\mathtt{f64} \to \mathtt{bool}$}
  }{
    \Env; \ranges{\Alg}{\forix~\iot{i}{n}}
      \vd {p~xs[i]} \To{Src} (\Env,~\ranges{\Alg}{\forix~\iot{i}{n}},~\ixfn{\iot{j}{1}}{\kw{true} => p[xs[i]]})
  }\lab{Unint}
  \\
  \Env(xs) = \ixfn{\iot{i}{n}}{\kw{true} => xs[i]}
}{
  \Env; \Alg \vd \Map{(\Fun{x}{p~x})}{xs}
  \To{Src}
  (\Env,~\Alg,~\ixfn{\iot{i}{n}}{\kw{true} => p[xs[i]]})
}\lab{Map}
\end{align*}
\end{small}%
To track positional dependencies backwards to the formal arguments of a function
declaration,
% ---strengthening pointful queries to the solver---
we substitute index functions into other index functions by reduction over
indexing symbols and variables.
Reduction contexts (\cref{fig:ctx}) define where a reduction is to
occur in an index function.
\begin{figure}
\begin{align*}
  \begin{array}{ccl}
  \Gsym  & ::=  & \Hole ~\mid~ g \gdelim \Gsym ~\mid~ \Gsym \gdelim g
  \\
  \Ksym  & ::=  & \Hole
                  ~\mid~ e + \Ksym
                  ~\mid~ e \cdot \Ksym
                  ~\mid~ x[\Ksym]
                  ~\mid~ \Sum{x}{\Ksym}{e}{s}
                  ~\mid~ \Sum{x}{e}{\Ksym}{s}
                  ~\mid~ \Sum{x}{e}{e}{\Ksym} \\
         & \mid & \neg \Ksym
                  ~\mid~ \Ksym \nd c
                  ~\mid~ c \nd \Ksym
                  ~\mid~ \Ksym \oo c
                  ~\mid~ c \oo \Ksym
                  ~\mid~ \Ksym \odot e
                  ~\mid~ e \odot \Ksym
  \end{array}
\end{align*}
\vspace{-3ex}
\caption{Reduction context grammar for guarded expressions ($\Gsym$) and symbols ($\Ksym$).}
\label{fig:ctx}
\end{figure}
$\G{c => e}$ denotes the guarded expression obtained by replacing
$\Hole$ with $c => e$ in the context $\Gsym$.  For instance, if
$\Gsym = (c_1 => e_1) \gdelim \Hole \gdelim (c_3 => e_3)$, then $\G{c_2 => e_2}$
is equivalent to $(c_1 => e_1) \gdelim (c_2 => e_2) \gdelim (c_3 => e_3)$.
% For convenience, we also allow eliminating $\Hole$;
% take $\G{}$ to mean $(c_1 => e_1) \gdelim (c_3 => e_3)$.
%
Similarly, $\K{e}$ denotes the symbol (or expression) obtained by replacing
$\Hole$ with $e$ in the context $\Ksym$.  For example, if
$\Ksym = e_1 + x_1[\Hole]$, then $\K{x_2}$ is $\Ksym = e_1 + x_1[x_2]$.

\begin{figure}
\input{deps/rewrite1.tex}
\vspace{-3ex}
\caption{Rewrite rules for index functions.}
\label{fig:substitute}
\end{figure}

Reductions are used in the
rewrite rules shown in \cref{fig:substitute}.
\textsc{I.Sub1} substitutes one index function into another
by combining their guarded expressions,
given that variables in the indexing expression, $e_0$, are free.
\textsc{E.Lift} treats scalars as one-element arrays,
enabling substitution.
\textsc{E.Sub1} substitutes an indexing symbol
anywhere given that the index function being indexed into
has only one guarded expression.
Via rules \textsc{I.1}, \textsc{I.2}, \textsc{G.1}, and \textsc{G.2},
this enables substitutions into domains and Sum symbols,
such as $\Sum{j}{e_1}{e_2}{x_1[j]}$ where $j$ is not free.
Note how in \textsc{G.2}, the truth of the guard is used to rewrite
its expression.
\textsc{E.Query$^\odot$} uses the solver to simplify symbols
beyond the scope of syntactical rewrites,
while \textsc{G.Falsify} uses the solver to falsify guards
by converting the guard to CNF and
then checking whether there exists a false conjunct
under assumption of all other conjuncts.
\textsc{G.Join-1} merges two guarded expressions
given that the value of the second guarded expression
simplifies to the value of the first guarded expression
under assumption of the first guard.
\textsc{G.Elim} eliminates guarded expressions
for which the guard is false.
For brevity, we omit analogous rules
\textsc{E.Query$^\land$} and \textsc{E.Query$^\lor$}
where $\odot$ in \textsc{E.Query$^\odot$}
is replaced by $\land$ and $\lor$, respectively;
\textsc{G.Join-2} where the roles of $c_0$ and $c_1$, and
$e_0$ and $e_1$, are swapped;
and rules for binary operations (similar to \textsc{If}).

Returning to \texttt{partition2}, we have,
via \textsc{Map}, \textsc{If}, \textsc{Const}, \textsc{E.Lift},
and \textsc{I.Sub1}:

\begin{centering}
\lstinline[mathescape=true]{flagsT = map (\c -> if c then 1 else 0) cs$\hsp|\hsp\ixfn{\iot{i}{n}}{p[xs[i]] => 1 \gdelim \neg p[xs[i]] => 0}$}\\
\end{centering}
\noindent Note how the positional relation between $xs$ and the values of $flagsT$ is
tracked automatically by substitution on $cs$.
In related systems, these must be established by the user (see
\cref{sec:rel-work}).

\label{sec:scan}
\begin{figure}
  \input{deps/convert2.tex}
  % \vspace{0.5em}
  \input{deps/convert3.tex}
  \vspace{-3.0ex}
  % \caption{Conversion and rewrite rules for $\kw{scan}$.}
  \caption{Additional rewrite rules as well as conversion of \texttt{scan} and \texttt{scatter}.}
  \label{fig:convert2}
\end{figure}

\emph{Scan: recurrences and sums.}  \Cref{fig:convert2} extends the rule set to
convert $\kw{scan}$ into an index function with a special recurrence symbol
$\Rec$ used for pattern matching in rewrites.
Rule \textsc{I.Sum} introduce sums by matching recurrences on linear polynomials
where
% with syntactical rewrite
\[
\textstyle
\mathrm{sum(n_1 \cdot c_1 + n_2 \cdot c_2 + \dots + n_v \cdot c_v,~a,~b)}
= n_1 \cdot \Sum{j}{a}{b}{c_1} + \dots + n_v \cdot \Sum{j}{a}{b}{c_v}.
\]
\textsc{G.Neg} rewrites negations on symbols
found outside guards into integer arithmetic.
\textsc{G.BoolToInt} flattens guarded expressions
% with exactly two guards and constants for values
% into a single guard,
matching user conversion from bool to integer.
% TODO should we maybe jsut restrict the user to
% use a prelude function bool2int if they want
% their proofs to go through? (Creating a rule for
% this function instead of ad-hoc trying to infer it.)
%
% (In both rules, type checking of the source program
% ensures that the integer value of the
% boolean symbol is one or zero.)
%
Both rules facilitate rewriting recurrences into
sums over boolean symbols. Converting $indicesT$ now yields:

\begin{centering}
\lstinline[mathescape=true]{indicesT = scan (\x y -> x + y) 0 flagsT$\hsp|\hsp\ixfn{\iot{i}{n}}{true => \Sum{j}{0}{i}{p[xs[j]]}}$}\\
\end{centering}

% \vspace{-0.5em}
\paragraph{Scatter: safety checks and segmented domains}
\label{sec:scatter}
\Cref{fig:convert2} converts $\kw{scatter}$ to index functions while verifying
safety; if safety cannot be shown, no rule will be matched
and the analysis terminates in failure.
The two conversion rules \textsc{Sc1} and \textsc{Sc2} first check safety
using \textsc{Ss1--3}.
\textsc{Ss1} verifies that the indices $is$ have no duplicates in the bounds of
the destination array $xs$, and \textsc{Ss2} and \textsc{Ss3} both verify
that the values in $vs$ do not depend on its indices (allowing for duplicate indices in $is$).
\textsc{Sc1} further verifies that the indices $is$, that fall
within the bounds of $xs$, are a permutation of the indices of $xs$.  The
resulting index function couples $is^{-1}$ to $is$, which is used in query
solving (e.g., \textsc{FPV2}).
\textsc{Sc2} verifies that $is$ is monotonically increasing,
producing an index function that has a domain with $e^m+1$ segments (per
\cref{sec:domains}). Out-of-bounds indices map to empty segments; this is
crucial, because the number of non-empty segments is generally statically
unknown.  The resulting index function is verbose, but simplifies via
the rewrite system.  For example, if $e_0$ rewrites to 0, the leading
$\ixfn{\iot{i}{e_0}}{(true => xs[i])}$ simplifies away (its domain is
empty)---this is the case when scattering ones into a flag array using an
exclusive scan of the shape array (\cref{fig:running-egs}). It also handles
scattering at an inclusive scan’s output, forming an initial linear segment. A
subsequent segmented scan with either flag array simplifies to the same index
function. %(See the supplementary material for segmented domain rules.)
A further simplification omits $c_k$ from the guards
if $c$ is implied by the segment being non-empty, checked via the queries
\[
\ranges{\Alg}{\forix~{\iot{i}{e^m}}} \vd e\{i \mapsto i+1\} - e > 0 \? c
\quad\text{and}\quad
\Alg \land i = e^m - 1 \vd e^n - e > 0 \? c.
\]

\begin{figure}
  \input{deps/substitute2.tex}
  \vspace{-5ex}
  \caption{Substitution rules for segmented domains.}
  \label{fig:substitute2}
\end{figure}
% XXX
% The remaining substitution rules are introduced
% after scatter (segmented domains).
%
Substitutions of segmented index functions into other index functions can be
reduced to \textsc{Sub1} by first transforming the index function
to have a linear domain using an $II$ array (or by treating $k$
as an uninterpreted function of the domain iterator $i$).
However, we define additional substitutions in \cref{fig:substitute2}
to automatically propagate structural information through index functions.
\textsc{Sub2} is similar to \textsc{Sub1}, but replaces the linear domain
of the target index function with the segmented domain
of the function being indexed into.
A query checks that $e_0$ is within bounds of segment $k$,
allowing the segmented domain to be adopted without modification.
\textsc{Sub3} further requires that the domains of two
segmented index functions are identical for substitution
to go through.
Finally, \textsc{Sub4} eliminates the iterator variable, $k$,
of the segmented index function being indexed into,
given unification of $e_k$ with $e_0$ yields a substitution
for $k$.
This is useful for scalar expressions that
index into segmented index functions,
in which case the domain, naturally, cannot be propagated.
If solving fails, $k$ can always be replaced by indexing
into an $II$ array corresponding to the segmented index function.
%
% XXX
% Finally, mention/add all the segmented modifications needed
% to previous rules. Such as:
% segmented versions of Map and Scan rules.
% segmented versions of rewrite rules (I.1,I.2 etc; check old paper).
%
Segmented versions of \textsc{Map}, \textsc{Scan}, \textsc{I.1}, and
\textsc{I.Sum} follow similar ideas and are ommited for brevity.
%\textsc{I.Sum} are straightforward and are therefore omitted; segmented
%\textsc{I.Sum} is in the supplementary material as an example.

\paragraph{Rewrite system and final remarks}
% XXX
% So far we've applied the right rules, at the right time, manually,
% define how rules recurse into eachother...
%
When converting the source language to index functions,
each conversion step is followed by rewrites applied to a fixed point:
\[
  \Env; \Alg \vd \esrc \To{Src} (\Env', \Alg', f),
  \hsp
  \Env; \Alg \vd f \leadsto (\Env'', \Alg'', f'),
\]
Rewrite rules are simply tried in the order that they appear here,
except for \textsc{I.Sum} and \textsc{I.Carry}, which
are only applied after \textsc{Let} as they introduce indexing with
bound variables and hence limit substitutions.
The source program is converted top-to-bottom;
top-level functions must be declared before using them.
This means all top-level functions have index functions
before they are applied. Thus application of function declarations
\begin{enumerate*}
  \item verifies that the actual arguments’ index functions satisfy
  the formal arguments’ pre-conditions;
  \item substitutes formal arguments with actual ones via the
  indexing reduction rules; and
  \item adds the previously verified post-condition to $\Alg$.
\end{enumerate*}

%\newpage

%%%%%%%%%%%%%%%%%%%%%%%%%%%%%%%%%%%%%%
%%% CHAPTER
%%%%%%%%%%%%%%%%%%%%%%%%%%%%%%%%%%%%%%
\section{Solving Queries}
\label{sec:query-solver}

The query solver has three main tasks: to solve equality and inequality queries
and to simplify internal expressions, e.g., enabling the side conditions related
to unification. \Cref{subsec:qs-lang-symtab} discusses the language of the query
solver and the construction of symbol tables such as the ones recording symbols'
ranges $Ranges$ and equivalences $Equivs$.  \Cref{subsec:solve-ineq}
presents our adaptation of the Fourier-Motzkin algorithm for solving
inequalities, and \cref{subsec:simplify} presents the simplification
engine that is aimed to support array indexing and sum symbols. While equality
queries are often satisfied through simplification,
\cref{subsec:eq-with-inj} presents a more advanced method that exploits
injective properties.

\begin{figure}
\begin{scriptsize}
  \begin{minipage}[t]{0.37\linewidth}
%  \vspace{-1ex}
    \[
    \begin{array}{l}
    \begin{array}{cclr}
    X & ::=  & x & \hfill\text{Non-Boolean Array}\\
         &~\mid~& DOR \ \ \overline{x} & \hfill\text{Disjoint Bool Arrays}\\
    \end{array}\\
    \begin{array}{cclr}
    s  & ::=  & x          & \text{Variable}\\
       & \mid & X[Poly^{ind}] & \text{Indexing}\\
       & \mid & \sum X[Poly^b : Poly^e] & \text{Slice Sum}\\
    \end{array}
    \smallskip\\
    \begin{array}{ccl}
    Term  & ::= & s^k ~\mid~ s^k ~*~ Term\\
    Poly  & ::= & k \ ~\mid~ Poly \ + \ k*Term
    \end{array}
    \end{array}
    \]\vspace{-4ex}\\
  \begin{center}
  \textcolor{blue}{\textsc{Example of Legal Ranges}}
  \end{center}
    \[
          \begin{array}{ccc}
          3 & \leq  n   \leq & \text{infty} \\
          n & \leq 3\cdot i \leq & \{5\cdot n, n^2\}\\
          i+n & \leq j \leq & i^2\\
          \{i+1, 5\} & \leq \sum X[i:i+j] \leq & j-1
          \end{array}
    \]\vspace{-4ex}\\
    \begin{center}
    \textcolor{red}{\textsc{Example of Illegal Ranges}}
    \end{center}
    \[
          \begin{array}{ccc}
          i & \leq n   \leq & i\cdot i \\
          n & \leq 2\cdot i \leq & 2\cdot n-5
          \end{array}
    \]

  \end{minipage}\begin{minipage}[t]{0.35\linewidth}
    \vspace{-3ex}
  \begin{algorithm}[H]
    \caption{\textsc{Solve}$^{\leq 0}_\Alg$}
    \label{alg:solve-leq0}
    \begin{algorithmic}[1]
    \Require a $Poly$ expression $\mathbf{p}$ in which monotonic indices have been translated to sums.
    \Ensure $\True$ if $\mathbf{p} \leq 0$ was verified, else $\textbf{false}$
    \State $\mathbf{p'}$ = \textsc{PeelOnRng}$_\Alg$(\textsc{Simplify}$_\Alg$ $\mathbf{p}$)
    \State \textbf{if} $\mathbf{p'}$ is constant $k$ \textbf{then return} $k \leq 0$
    \State $\mathbf{s}$  = \textsc{FindSym}$_\Alg$ $\mathbf{p'}$
    \State $(\mathbf{lbs}, \mathbf{k}, \mathbf{ubs})$ = \textsc{getRange}$_\Alg$ $\mathbf{s}$
    \State \textbf{rewrite} $\mathbf{p'}$ as $\mathbf{p}_a \cdot \mathbf{s} + \mathbf{p}_b$
    \For{\textbf{each} $(\mathbf{l},\mathbf{u}) \in \mathbf{lbs} \times \mathbf{ubs}$}
        \State ($\mathbf{p1},\mathbf{p2})$ = ($\mathbf{l} \cdot \mathbf{p_a} + \mathbf{k}\cdot \mathbf{p_b}$, $\mathbf{u} \cdot \mathbf{p_a} + \mathbf{k}\cdot \mathbf{p_b}$)
%        \State success1 = ( \textsc{Solve}$^{\leq 0}_\Alg$ $\mathbf{p}_1$ ) $\wedge$\\
%                          ( \textsc{Solve}$^{\leq 0}_\Alg$ $\mathbf{lb} \cdot \mathbf{p}_a + k\cdot p_b$ )
%        \State success2 = ( \textsc{Solve}$^{\leq 0}_\Alg$ $0 - \mathbf{p}_a$ ) $\wedge$\\
%                          ( \textsc{Solve}$^{\leq 0}_\Alg$ $\mathbf{ub} \cdot \mathbf{p}_a + k\cdot p_b$ )
%                           %l*p_1 + k*p_2
%        \If{sucess1 $\vee$ success2}
        \If{( \textsc{Solve}$^{\leq 0}_\Alg \ (- \mathbf{p}_a)$ $\wedge$ \textsc{Solve}$^{\leq 0}_\Alg \ \mathbf{p2}$ ) $\vee$
        \\\hspace{6ex}( \textsc{Solve}$^{\leq 0}_\Alg \ \mathbf{p}_a$ $\wedge$ \textsc{Solve}$^{\leq 0}_\Alg \ \mathbf{p1}$ )}
          \State \Return $\True$
        \EndIf
    \EndFor
    \State \Return $\textbf{false}$
\end{algorithmic}
\end{algorithm}
  \end{minipage}\begin{minipage}[t]{0.25\linewidth}
    \vspace{-3ex}
  \begin{algorithm}[H]
    \caption{\textsc{Simplify}$_\Alg$}
    \label{alg:simplify}
    \begin{algorithmic}[1]
    \Require a $Poly$ expression
    \Ensure a $Poly$ expression semantically equivalent with the input
    \While{Fix Point Not Reached}
    \State Apply \textsc{SubstEquivs}$_\Alg$
    \State Apply \textsc{0Sum}
    \State Apply to a fix point \textsc{UnBef}
    \State Apply to a fix point \textsc{B1-5}
    \State Apply \textsc{0Sum}
    \State Apply to a fix point \textsc{UnAft1-5}
    \EndWhile
    \State \Return  fix-point expression
\end{algorithmic}
\end{algorithm}
%
%  \begin{algorithm}[H]
%    \caption{\textsc{findSym}$_\Alg$}
%    \label{alg:find-sym-to-elim}
%    \begin{algorithmic}[1]
%    \Require a $Poly$ expression $\mathbf{p}$
%    \Ensure the symbol $\mathbf{sym}$ whose ranges depend on the most other symbols
%    \State $\mathbf{len}$ = 0
%    \State $\mathbf{sym}$ = \textbf{null}
%    \For{\textbf{each} symbol $\mathbf{s}$ of $\mathbf{p}$}
%        \State $\mathbf{tc}$ = compute the transitive closure of
%                    symbols appearing in the ranges starting with
%                    those of $\mathbf{s}$
%        \If{$\mathbf{len} \ \ \leq \ \ \text{length}~\mathbf{tc}$}
%          \State $\mathbf{sym}$ = $\mathbf{s}$
%          \State $\mathbf{len}$ = length $\mathbf{tc} $
%        \EndIf
%    \EndFor
%    \State \Return $\mathbf{sym}$
%\end{algorithmic}
%\end{algorithm}
  \end{minipage}
\end{scriptsize}
\vspace{-1ex}
\caption{Grammar for symbols ($s$), polynomial ($Poly$) and guarded expressions ($g$), and index functions $ixfn$.}
\label{fig:solver-grammar}
\end{figure}

\subsection{Query Solver Language Lowering and Symbol Tables}
\label{subsec:qs-lang-symtab}

The left side of~\cref{fig:solver-grammar} shows the language of the query solver.
Capital $X$ denotes either the name of an integral array $x$ or a disjunction of a
sequence of disjoint/orthogonal boolean arrays, denoted $\text{DOR} \ \overline{x}^v$,
i.e., for any legal index $i$, $(\text{DOR} \ \overline{x}^v)[i] = x_1[i] \vee \ldots \vee x_{v}[i]$
{\em and} there is at most one $h\in 0\ldots v$ holding a true value. A singleton
sequence is always legal, and an empty sequence results in false.
%, the neutral element of logical or.
Computationally, boolean values are treated as integers, $1$ for true and $0$ for false.

The solver uses a polynomial representation similar to the one of index
functions, denoted $Poly$, except that its symbols are restricted to scalar
variables $x$, indexing and sums of array slices (of unit stride)
over $X$.
%
%\footnote{
%Generalization to non-one strides is possible.
%}.
%
High-level queries are lowered to $Poly$ in the standard way, by
binding (bilaterally) untranslatable expressions to fresh variables,
whose properties are documented, as much as possible, in other symbol tables.
For example, the guards $\overline{c}^v$ of an index function give raise
to $v$ fresh symbols $\overline{x}^v$, where each one of them is bound
in the $\Alg.DOR$ symbol table to the whole sequence $\overline{x^v}$.
While the exact meaning of the predicates is lost, the solver can still
conduct reasoning based on the property that at any index, exactly one
of $\overline{x}^v$ is true, which subsumes disjointness.

Similarly, for the purpose of inequality solving, a monotonically-increasing
array $x$ of length $n$ is translated to the sum of a slice of a fresh
array variable $a$, whose elements are positive. Assuming that $x$'s elements
are known to be within $[e^{lb},e^{ub}]$, then a valid
index $x[i]$ is translated to $e^{lb} + \sum a[0:i]$, while encoding
in $Ranges$ that $a$ has positive elements (and
$\sum a[0:n-1] \leq e^{ub} - e^{lb}$). Strictly-increasing monotonicity is
similarly handled by means of a strictly positive array.

%The range and equivalence symbol tables, denoted $Ranges$ and $Equiv$ are
The $Ranges$ and $Equiv$ symbol tables are constructed to conform with a
``triangular'' shape that essentially requires that an ordering of the bound
symbols exists, such that the range or equivalent rewrite of a symbol may
only depend on its predecessors.
Bindings $sym \mapsto bnd$ are added by processing boolean expressions denoting
(in)equalities, e.g., corresponding to code properties (e.g., branch conditions
or loop counters) or query premises. From them, a set of candidate bindings
are identified and the invalid ones are rejected, whenever it is found that
the leading name of $sym$ appears in the transitive closure of variable names
through $Ranges$ and $Equiv$ that starts from $bnd$.
If there are several legal candidates for the same (in)equality, the winning
symbol is selected by a set of heuristics, e.g., the one appearing latest in
program order.
For example, assuming an empty $Equiv$, the binding $i \mapsto B[i]$ is illegal
because $i\in \{i,B\}$, but $B[i] \mapsto i$ is legal because the leading
symbol $B \not\in \{i\}$. Examples of legal and illegal $Ranges$ tables are
also shown in \cref{fig:solver-grammar}.

\subsection{Solving Inequalities}
\label{subsec:solve-ineq}

Inequalities are reduced to form $p \leq 0$ and solved by
the adaptation of Fourier-Motzkin elimination~\cite{Fourier-orig,Williams01111986}
presented in \cref{alg:solve-leq0}, which proceeds by simplifying $p$,
which is necessary in the presence of index and sum symbols,
%by performing the equivalent rewrites of $Equiv$ and then simplifying the result.
%The later step is necessary in the presence of index and sum symbols,
otherwise the canonical polynomial representation $Poly$ is optimal.
The next line implements the base case, i.e., when the simplified
expression $p'$ is a value $k$.
If not, \textsc{FindSym$_\Alg$} determines the next symbol $s$ to
eliminate, as the symbol of $p'$ whose range transitively depends
on the largest number of (other) distinct symbols.
The range of $s$, computed by \textsc{GetRange$_\Alg$}, takes the form
$(\overline{lbs}, k, \overline{ubs})$, where $k > 0$ is a constant and
$\overline{lbs}$ and $\overline{ubs}$ are expression sequences having
the semantics
$\kw{max}(\overline{lbs}) ~\leq~ k\cdot s ~\leq~ \kw{min}(\overline{ubs})$.
\textsc{GetRange$_\Alg$} looks up the ranges recorded in $Ranges$, e.g.,
an index $x[i]$ may possibly have several lower/upper bounds originating
in program branches or query premises, which are further expanded with
the range of $x$'s elements, if $x$ has one. As well, \textsc{GetRange$_\Alg$}
infers ranges, e.g., if array $x$ is strictly positive,
then a lower bound for $\sum x[lb : ub]$ is $ub - lb + 1$, if the
later can be proven positive, and similar for the upper bound of
$\sum (\text{DOR} \ \overline{x})[lb:ub]$.

Finally, $p' \leq 0$ is re-written as $p_a\cdot (k\cdot s) + k\cdot p_b \leq 0$,
such that $s \not\in p_b$, and $k\cdot s$ is conservatively replaced---i.e.,
with its upper bound if $p_a \geq 0$ and with its lower bound otherwise---generating
two sub-problems for each combination of lower and upper bounds from the
sequence. If any succeeds the query has been verified, otherwise the result
is unknown. Our adaptation allows polynomial ranges and polynomial queries,
as well; for example, assuming the valid $Ranges$ in \cref{fig:solver-grammar},
the query $n^2 + 3\cdot n - j^2 - j \leq 0$ succeeds. The first eliminated symbol is
$j$, since its ranges transitively depend on both $i$ and $n$, and the next one
is $i$. Note that if $n$ is chosen as the first symbol to eliminate,
the query fails, immediately resulting in $\infty \leq 0$.
\textsc{Solve}$^{\leq 0}_\Alg$ is guaranteed to terminate because
(1) the ``triangular'' shape of $Ranges$ and $Equivs$ do not introduce
cycles, and (2) at every step, the target symbol will be eliminated in
a number of steps equal to its polynomial degree, which is finite.

%
% shouldn't the substitutions be performed in simplify?

\subsection{Simplifying Index and Sum-of-Slice Expressions}
\label{subsec:simplify}

We motivate our technique on one of the simpler queries generated by
{\tt partition2},
%(see~\cref{fig:running-egs}),
namely:
$j + \sum C[j+1 : n-1] - \sum C[0:i-1] > 0$,
where the context is
$Ranges = \{~0\leq n \wedge 0 \leq i \leq n-1 \wedge 0 \leq j \leq i-1 \}$
and $Equivs = \{ C[i] = 1 \wedge C[j] = 0 \}$ and
$C = \text{DOR} \ c$ has elements in the implicit range $[0,1]$.
Note that (1) $Equivs$ cannot be applied because $C[i]$ and $C[j]$
do not appear in the sums, and (2) the sum subtraction cannot be
simplified because it cannot be proven that the two slices overlap.
Applying~\cref{alg:solve-leq0} will end in a clearly false
subquery $j > i$.
%
%As such, the expression seems unsimplifiable.
%Applying~\cref{alg:solve-leq0} on it will
%first replace symbol\footnote{
%$\sum C[j+1 : n\text{-}1]$ has upper bound $n\text{-}j\text{-}1$,
%resulting in transitive
%closure $\{i,j,n\}$, hence it is the most ``dependent'' symbol.
%} $\sum C[j+1 : n-1]$ with its lower bound
%$0\cdot (n-1-j-1+1) = 0$, then it will replace symbol $\sum C[0:i-1]$ with
%its upper bound $i-1+1$, resulting in query $j > i$, which is actually
%false since our context states $j < i$.

Our strategy is organized in three steps: The first is to extend the
ends of the summed slices with the indices that have bindings in $Equiv$.
This enables the second step, which perform simplifications across
sum-sum or sum-index pairs. Finally, the indices that have a binding
in $Equivs$ are separated from sums and substituted with their rewrite.
With our example, the first step will result in
$j + \sum C[j : n-1] - \sum C[0:i] + 1 > 0$. Since both slices are
now provably overlapping, i.e., context says $i > j$, the common part
can be eliminated, resulting in $j + \sum C[i+1:n-1] - \sum C[0:j-1] + 1 > 0$.
Applying~\cref{alg:solve-leq0} will first replace\footnote{
$\sum C[0 : j-1]$ has upper bound $j$, resulting in transitive
closure $\{i,j,n\}$, hence it is the most ``dependent'' symbol.
}
$\sum C[0:j-1]$ with its upper bound $j$ then will replace
$\sum C[i+1:n-1]$ with its lower bound $0$, resulting in
$j - j + 1 > 0$, which succeeds after simplification.

\begin{figure}
  \sempart{Unary Simplification $\To{UnS}$ uses helper $\To{Eqv}$ to rewrite symbol $s$ as Poly-expression $p$.}{\Alg \vd s \To{UnS} p}
    \begin{scriptsize}\vspace{-2ex}\\
    \begin{minipage}[t]{0.24\linewidth} % 0.22 is enough
    \vspace{3ex}
    \begin{align*}
    \inference{
      \Alg.Equiv(s) = p
    }{
      \Alg \vd s \To{Ev} p
    }
    \tagsc{Eqv1}
    \end{align*}
    \end{minipage}\begin{minipage}[t]{0.39\linewidth}
    \begin{align*}
    \inference{
      s = (\text{DOR} ~\overline{x}^v)[p^{ind}]
      \\
      \exists i\in 1\ldots v~\text{|}~\Alg.Equiv(x_i[p^{ind}]) = 1
    }{
      \Alg \vd s \To{Eqv} 1
    }
    \tagsc{Eqv2}
    \end{align*}
    \end{minipage}
    \begin{minipage}[t]{0.28\linewidth} % 0.26 is enough
    \vspace{-1ex}
    \begin{align*}
    \inference{
      s = \sum x[p^{b} : p^{e}]
      \\
      \colorbox{black!10}{$\Alg \vd \True \? p^{b} > p^{e}$}
    }{
       \Alg \vd s \ \To{UnS} \ 0
    }
    \tagsc{0Sum}
    \end{align*}
    \end{minipage}\\
    \begin{minipage}[t]{0.47\linewidth}
    \vspace{-4ex}
    \begin{align*}
    \inference{
      s = \sum X[p^{b} ~:~p^{e}]
        \hsp
        \colorbox{black!10}{$\Alg \vd \True \? p^{e} + 1 \geq p^{b}$}
      \\
      \Alg \vd X[p^{b}-1] \To{Eqv} p^x_{b}
        \hsp s' = \sum x[p^{b}-1:p^{e}]
    }{
      \Alg \vd s \ \To{unS} \ s' - p^x_{b}
    }
    \tagsc{UnBef}
    \end{align*}
    \end{minipage}
    \begin{minipage}[t]{0.30\linewidth}
    \vspace{-3ex}
    \begin{align*}
    \inference{
      s = \sum x[p^{b} : p^{e}]
      \\
      \colorbox{black!10}{$\Alg \vd \True \? p^{b} = p^{e}$}
    }{
       \Alg \vd s \ \To{UnS} \ X[p^{b}]
    }
    \tagsc{UnAft1}
    \end{align*}
    \end{minipage}
    \begin{minipage}[t]{0.22\linewidth}
    \vspace{1ex}
    \begin{align*}
    \inference{
      \Alg \vd s \To{Eqv} p
    }{
       \Alg \vd s \ \To{UnS} \ p
    }
    \tagsc{UnAft2}
    \end{align*}
    \end{minipage}
    \\
    \begin{minipage}[t]{0.35\linewidth}
    \vspace{-1ex}
    \begin{align*}
    \inference{
      s = \sum X[p^{b} ~:~p^{e} ]
        \hsp
        \colorbox{black!10}{$\Alg \vd \True \? p^{e} \geq p^{b}$}
      \\
      \Alg \vd X[p^{b}] \To{Eqv} p^x_{b}
        \hsp s' = \sum x[p^{b}+1 : p^{e}]
    }{
      \Alg \vd s \ \To{UnS} \ s' + p^x_b
    }
    \tagsc{UnAft3}
    \end{align*}
    \end{minipage}%\hfill
%
%    \begin{minipage}[t]{0.38\linewidth}
%    \vspace{-3ex}
%    \begin{align*}
%    \inference{
%      s = \sum (\text{DOR}~\overline{x}^v)[p^b~:~p^e]
%      \\
%        \colorbox{black!10}{$\Alg \vd \True \? p_u+1\geq p^b$}
%      \hsp v > 0
%      \\
%      \Alg.DOR(x_1) = \overline{y}^w
%      \hsp w = v-1
%    }{
%       \Alg \vd s \ \To{UnS} \ p^e + 1 - p^b
%    }
%    \tagsc{UnAft4}
%    \end{align*}
%    \end{minipage}\\
%
%    \begin{minipage}[t]{0.29\linewidth}
%    \begin{align*}
%    \inference{
%      s = (\text{DOR}~\overline{x}^v)[p] \hsp v > 0
%      \\
%      \Alg.DOR(x_1) = \overline{y}^w
%      \hsp w = v-1
%    }{
%       \Alg \vd s \ \To{UnS} \ 1
%    }
%    \tagsc{UnAft5}
%    \end{align*}
%    \end{minipage}
%
    \begin{minipage}[t]{0.25\linewidth}
    \vspace{2ex}
    \begin{align*}
    \inference{
      s = \sum X [\_:\_]
      \\ \textbf{OR} \ \ s = X[\_]
      \\
      X = \text{DOR}~\overline{x}^v
        \hsp
        v=0
    }{
       \Alg \vd s \ \To{UnS} \ 0
    }
    \tagsc{UnAft4}
    \end{align*}
    \end{minipage}\hfill
    \begin{minipage}[t]{0.4\linewidth}
    \vspace{-1ex}
    \begin{align*}
    \inference{
      s = \sum X[p^{b} ~:~p^{e} ]
        \hsp
        \colorbox{black!10}{$\Alg \vd \True \? p^{e} \geq p^{b}$}
      \\
      \Alg \vd X[p^{e}] \To{Rng} \_
        \hsp
        s' = \sum x[p^{b} : p^{e}-1]
    }{
      \Alg \vd s \ \To{UnS} \ s' + X[p^e]
    }
    \tagsc{PeelOnRng}
    \end{align*}
    \end{minipage}
    \end{scriptsize}
  \sempart{Rewrites a sum of two terms $k_1\cdot s_1\cdot t_1 + k_2\cdot s_2\cdot t_2$ as a Poly $p$.}{\Alg \vd ((k_1,s_1,t_1),(k_2,s_2,t_2)) \To{BinS} p}
    \begin{scriptsize}
    \begin{minipage}[t]{0.53\linewidth}
    \vspace{-1ex}
    %\hspace{-4ex}
    \begin{align*}
    \inference{
      s_1 = \sum X_1[p^b_1~:~p^e_1] \hsp s_2 = \sum X_2[p^b_2~:~p^e_2]
      \\
      X_1 = X_2 \hsp k_1 = k_2 \hsp t_1 = t_2
        \hsp \colorbox{black!10}{$\Alg \vd \True \? p^e_1 + 1 = p^b_2$}
      \\
      (\colorbox{black!10}{$\Alg \vd \True \? p_1^e + 1 \geq p_1^b$}
        \ \textbf{OR}
        \ \colorbox{black!10}{$\Alg \vd \True \? p_2^e + 1 \geq p_2^b$})
    }{
      \Alg \vd ((k_1,s_1,t_1),(k_2,s_2,t_2)) \To{BinS} k_1 \cdot t_1 \cdot \sum X_1[p_1^{b} : p_2^{e}]
    }
    \tagsc{B1}
    \end{align*}
    \end{minipage}
    \begin{minipage}[t]{0.46\linewidth}
    \begin{align*}
    \inference{
      s_1 = \sum X_1[p^b~:~p^e] \hsp s_2 = X_2[p^i] \hsp X_1 = X_2
      \\
      k_1 = k_2 \hsp t_1 = t_2
        \hsp
        \colorbox{black!10}{$\Alg \vd \True \? p^e + 1 \geq p^b$}
      \\
      \colorbox{black!10}{$\Alg \vd \True \? p^i = p^e + 1$}
        \
        s' = \sum X_1[p^b~:~p^e+1]
    }{
      \Alg \vd ((k_1,s_1,t_1),(k_2,s_2,t_2)) \To{BinS} k_1 \cdot s' \cdot t_1
    }
    \tagsc{B2}
    \end{align*}
    \end{minipage}\medskip\\

    \begin{minipage}[t]{0.99\linewidth}
    \begin{align*}
    \inference{
      k_1 = k_2 \ \ t_1 = t_2 \ \ s_1 = \sum (\text{DOR} \ \ \overline{x}^{v})[p_x^b : p_x^e]
        \hsp s_2 = \sum (\text{DOR} \ \ \overline{y}^{w})[p_y^b : p_y^e]
        \hsp y_1 \in \Alg.DOR(x_1)
        \ \ \overline{x} \cap \overline{y} = \emptyset
        \ \ \overline{z} = \overline{x} \cup \overline{y}
      \\
%      s_1' = \sum (\text{DOR}~\overline{x})[p_x^b : p_y^b-1]
%      \ \
      v > 0
      \hsp w > 0
      \hsp
      % (s'_2, s_3') =
      %   \text{$\begin{cases}
      %             (\sum (\text{DOR}~\overline{z})[p_y^b : p_x^e],
      %             ~\sum (\text{DOR}~\overline{y})[p_x^e+1 : p_y^e])
      %                & \textbf{if} \
      %                   \colorbox{black!10}{$\Alg \vd \True \? p_x^b \leq p_y^b \leq p_x^e \leq p_y^e$}\\
      %             (\sum (\text{DOR}~\overline{z})[p_y^b : p_y^e],
      %             ~\sum (\text{DOR}~\overline{y})[p_y^e+1 : p_x^e])
      %                & \textbf{if} \
      %                   \colorbox{black!10}{$\Alg \vd \True \? p_x^b \leq p_y^b \leq p_y^e \leq p_x^e$}
      %          \end{cases}$}
      (s'_2, s_3') =
                  (\sum (\text{DOR}~\overline{z})[p_y^b : p_x^e],
                  ~\sum (\text{DOR}~\overline{y})[p_x^e+1 : p_y^e])
      \hsp
      \text{$\colorbox{black!10}{$\Alg \vd \True \? p_x^b \leq p_y^b \leq p_x^e \leq p_y^e$}$}
    }{
      \Alg \vd ((k_1,s_1,t_1),(k_2,s_2,t_2)) \To{BinS}
                k_1 \cdot t_1 \cdot \sum (\text{DOR}~\overline{x})[p_x^b : p_y^b-1] \ + \
                k_1 \cdot s_2' \cdot t_1 \ + \ k_1 \cdot s_3' \cdot t_1
    }
    \tagsc{B3}
    \end{align*}
    \end{minipage}\medskip\\
    \begin{minipage}[t]{0.77\linewidth}
    \begin{align*}
    \inference{
      k_1 = -k_2 \hsp t_1 = t_2 \hsp s_1 = \sum X[p_x^b~:~p_x^e] \hsp s_2 = \sum Y[p_y^b~:~p_y^e] \hsp X = Y
      \\
      % (s'_1, s'_2) =
      %   \text{$\begin{cases}
      %             (\sum X[p_x^b : p_y^b-1], ~\sum X[p_x^e+1 : p_y^e])
      %                & \textbf{if} \
      %                  \colorbox{black!10}{$\Alg \vd \True \? p_x^b \leq p_y^b \leq p_x^e \leq p_y^e$}\\
      %             (\sum X[p_x^b : p_y^b-1], ~\sum X[p_y^e+1 : p_x^e])
      %                & \textbf{if} \
      %                  \colorbox{black!10}{$\Alg \vd \True \? p_x^b \leq p_y^b \leq p_y^e \leq p_x^e$}
      %          \end{cases}$}
      (s'_1, s'_2) = (\sum X[p_x^b : p_y^b-1], ~\sum X[p_y^e+1 : p_x^e])
      \hsp
      \text{$\colorbox{black!10}{$\Alg \vd \True \? p_x^b \leq p_y^b \leq p_y^e \leq p_x^e$}$}
    }{
      \Alg \vd ((k_1,s_1,t_1),(k_2,s_2,t_2)) \To{BinS} k_1 \cdot s_1' \cdot t_1 + k_2 \cdot s_2' \cdot t_2
    }
    \tagsc{B4}
    \end{align*}
    \end{minipage}\medskip\\
    \begin{minipage}[t]{1.0\linewidth}
    \begin{align*}
    \inference{
      k_1 = -k_2 \hsp t_1 = t_2 \hsp s_1 = \sum (\text{DOR} \ \ \overline{x})[p_x^b : p_x^e]
        \hsp s_2 = \sum (\text{DOR} \ \ \overline{y})[p_y^b : p_y^e]
        \hsp \overline{z} = \overline{x} \cap \overline{y}
        \hsp \overline{z} \not= \emptyset
        \hsp \overline{x'} = \overline{x} - \overline{w}
        \hsp \overline{y'} = \overline{y} - \overline{w}
      \\
        (y_1 = x_1 \ \ \textbf{OR} \ \ y_1 \in \Alg.DOR(x_1))
        \hsp Z = \text{DPR} \ \ \overline{z}
        \hsp \Alg \vd ((k_1,~Z[p_x^b:p_x^e],t_1),~(k_2,~Z[p_y^b : p_y^e],~t_2)) \To{BinS} p^z
    }{
      \Alg \vd ((k_1,s_1,t_1),(k_2,s_2,t_2)) \To{BinS}
        k_1 \cdot t_1 \cdot \sum (\text{DOR} \ \ \overline{x'})[p_x^b : p_x^e] \ + \
        k_2 \cdot t_2 \cdot \sum (\text{DOR} \ \ \overline{y'})[p_y^b : p_y^e] \ + \ p^z
    }
    \tagsc{B5}
    \end{align*}
    \end{minipage}
    \end{scriptsize}
  \vspace{-2ex}
  \caption{Rules for Unary and Binary Simplification of Expressions in Polynomial Representation.}
  \label{fig:unary-bin-simp}
\end{figure}

Figure~\ref{fig:unary-bin-simp} shows the inference rules of the simplification
engine, which are composed as depicted in~\cref{alg:simplify}
(\textsc{Simplify}$_\Alg$).
Unary simplifications replace a symbol with an equivalent expression.
Helper rules \textsc{Eqv1} and \textsc{Eqv2} replace a symbol with its binding in
$Equiv$, and a $\text{DOR}$ index with $1$ if one of the arrays in the disjunction
has $1$ as the binding of that index. \textsc{0Sum} replaces sums of provably
empty slices with $0$. \textsc{UnBef} extends a slice sum with an ending index
that is bound in $Equivs$. It requires that the slice is provably potentially
empty in a nice way $PENW$, i.e., its lower bound is at most $1$ higher than
its upper bound, which guarantees that the extended slice contains said element.
% is non-empty slice and
%
\textsc{Simplify}'s first stage consists of applying the equivalences in
$Equivs$ (\textsc{SubstEquivs}$\Alg$), \textsc{0Sum}, and a fixed-point
application of \textsc{UnBef} on any matching symbols of degree one.

The second stage consists of applying the binary rules \textsc{B1-5} to
a fix point. They denote an equivalent rewrite $p$ across any (sum of) two terms
$k_{1,2}\cdot s_{1,2}\cdot t_{1,2}$ of a polynomial $p^{tgt}$,
where $k_{1,2}$ are constants, $s_{1,2}$ are symbols of degree $1$,
and $t_{1,2}$ are a product of symbols not containing $s_{1,2}$.

The rewrite of $p^{tgt}$ is thus
$p^{tgt} - k_1\cdot s_1\cdot t_1 - k_2\cdot s_2\cdot t_2 + p$.
%The result $p$
%is equivalent with the sum of the two terms and $p^0$ is re-written as
%$p^0 - k_1\cdot s_1\cdot t_1 - k_2\cdot s_2\cdot t_2 + p$.
%
\textsc{B1} rewrites two sums of slices that are in continuation of each
other as one sum, and requires that at least one of them is $PENW$.
\textsc{B2} extends a slice sum with an end index.
\textsc{B3} simplifies two sums of overlapping slices corresponding to
$\text{DOR}$ arrays, whose sequences of variable names do not overlap,
but are in the same $\Alg.DOR$ class. \textsc{B4} eliminates the common
part of two overlapping slice sums, corresponding to the same array, that
are subtracted, and \textsc{B5} extends this simplification of slice-sum
subtraction to $\text{DOR}$ arrays.
Two additional rules (not shown), analogous to \textsc{B3} and \textsc{B4},
handle the cases when $p_x^b \leq p_y^b \leq p_y^e \leq p_x^e$
and $p_x^b \leq p_y^b \leq p_x^e \leq p_y^e$, respectively.
%(The rules are given in the supplementary material.)

The final pass applies \textsc{0Sum} followed by a fix-point
application of unary rules \textsc{UnAft1-6}. They collapse a
singleton slice to an index (\textsc{UnAft1}), replace symbols
bound in $Equivs$ (\textsc{UnAft2}), peel-off an index bound
in $Equivs$ from an end of a slice sum (\textsc{UnAft3}), and
eliminate null \text{DOR} slices/indices. Finally, \textsc{PeelOnRng},
not part of simplification, peels off an end-of-slice-sum index
that has a more specialized range than the one of the whole
array, which improves the accuracy of \textsc{Solve}$_\Alg$.

\subsection{Exploiting Injective Properties with the Equality Solver}
\label{subsec:eq-with-inj}

Queries $\Alg \vd c_1 \? c_2$ are discharged by the equality solver through
syntactical rewrites on the internal language
(\cref{fig:internal-grammar}). First the antecedent of the query, $c_1$, is
expanded with its transitive equalities.  Then, the transitive equalities are
rewritten using properties found in $\Alg$. This expanded query is
sent to the inequality solver (which may solve $s_1 == s_2$ as the queries $s_1 \geq s_2 \land s_1 \leq s_2$.)

More specifically, $c_1$ is converted to CNF yielding $c_1'$, and the first conjunct
is chosen to be the \emph{guide equation}. The right hand side of the guide equation
is then substituted for the left hand side everywhere in $c_1'$.
Transitive equalities are then extracted by treating each equality as edges in a graph
(symbols in $c_1$ being the nodes) and then computing a spanning forest using depth-first search.
Finally, transitive equalities are conjoined with $c_1'$ and
the syntactical rewrites are applied. For example, a rule exploiting injectivity is:\vspace{-1ex}
\[
\small
\inference{
  \RCD = \Alg.Inj(x)
  \hsp
  \text{\colorbox{black!6}{
    $\Alg \vd \kw{true} \? i \in \RCD \land j \in \RCD$
  }}
}{
  \Alg \vd \K{x[i] = x[j]} \to i = j
}\textsc{(Eq.Inj)}
\]
In rules such as \textsc{InjGe} the appropriate guide
is obvious (equality on the values of the guarded expressions); if there is
no obvious choice, the guides can be tried exhaustively.
%several queries can be dispatched, each with a different guide.

\newpage
%%%%%%%%%%%%%%%%%%%%%%%%%%%%%%%%%%%
%%% EVALUATION
%%%%%%%%%%%%%%%%%%%%%%%%%%%%%%%%%%%

\begin{figure}
\begin{minipage}[t]{0.48\linewidth}
\begin{lstlisting}[language=Futhark,basicstyle=\scriptsize,frame=none]
def filter_by [n] 't (cs: [n]bool) (xs: [n]t)
  : []t | \ys -> FiltPart ys xs (\i -> cs[i]) (\_ -> true) =
  @\textrm{\emph{Analogous to partition2 in \cref{fig:running-egs}; uses \textbf{scatter}}.}@

def get_smallest_pairs [n]
    (n_verts: i64) (n_es: i64)
    (es: [n]i64 | Range es (0, n_verts))
    (is: [n]i64 | Inj is (@$-\infty$@,@$\infty$@))
    : ([]i64, []i64) | \(es', is') ->
         Inj es' (@$-\infty$@,@$\infty$@) && Inj is' (@$-\infty$@,@$\infty$@) =
  let H = hist i64.min n_es n_verts es is
  let cs = map2 (\i j -> H[i] == j) es is
  let xs = filter_by cs es
  let ys = filter_by cs is
  in (xs, ys)
\end{lstlisting}
\end{minipage}
\vline\hspace{0.2em}
\begin{minipage}[t]{0.48\linewidth}
\begin{lstlisting}[language=Futhark,basicstyle=\scriptsize]
def kmeans_ker [num_cols] [nnz] [n]
    (row: i64 | Range row (0,n))
    (pointers: [n+1]i64 | Range pointers (0,nnz))
    (cluster: [num_cols]f32) (values: [nnz]f32)
    (indices: [nnz]i64 | Range indices (0,num_cols))
    : f32 =
  let index_start = pointers[row]
  let nnz_sgm = pointers[row+1] - index_start
  in loop (correction) = (0) for j < nnz_sgm do
      let element_value = values[index_start+j]
      let column = indices[index_start+j]
      let cluster_value = cluster[column]
      let diff = element_value - 2 * cluster_value
      let res = correction + diff*element_value
      in  res
\end{lstlisting}
\end{minipage}\\
\vspace{-0.4em}
\caption{Compute kernels from Maximal Matching (left)
and sparse $k$-means (right).}
\label{fig:eval}
\end{figure}

\section{Evaluation}
\label{sec:eval}
We illustrate the practicality of our system
through three case studies.
\vspace{0.2em}

\emph{Statically safe scatters.}
We verify Futhark's maximal matching benchmark,\footnote{
  Found at \url{https://github.com/diku-dk/futhark-benchmarks}.
} which implements a graph algorithm
from the Problem Based Benchmark Suite \cite{PBBS}
that iteratively filters graph edges using $\Scattertt$ operations.
The most challenging kernel to verify, \lstinline{get_smallest_pairs}, is shown
in \cref{fig:eval}.  The postcondition says that the values in each output array
\lstinline{es'} and \lstinline{is'} are unique.  We automatically verify both of
these properties using the injectivity of \lstinline{is}.
% \footnote{
%   The \lstinline{Range} property on \lstinline{es} is used only for bounds checking.
% }
After index function inference, we have $\Env(es) = \ixfn{\iot{i}{n}}{true
=> es[i]}$ and $\Alg.Inj(is) = (-\infty, \infty)$ in the environment. The
postcondition \lstinline[mathescape]{Inj es' ($-\infty$,$\infty$)} starts a
proof query, which will match first \textsc{InjV2} and then \textsc{InjF1}.
\textsc{InjF1} attempts two rewrites matching \textsc{InjGE} and \textsc{MonGe},
respectively. \textsc{InjGe} succeeds:
% but the latter fails.
% We illustrate the successful derivation below.
%
\begin{footnotesize}
% {\fontsize{7.5pt}{9pt}\selectfont
\begin{align*}
  \inference{
    \inference{
      \inference{
        \text{\colorbox{black!6}{
          $\Alg' \vd
            % e_h = \sigma(e_l)
            \textcolor{blue}{es[i]} = \textcolor{magenta}{es[j]}
              \land H[is[\textcolor{blue}{es[i]}]] = is[i]
              \land H[is[\textcolor{magenta}{es[j]}]] = is[j]
              \? i=j$}
        }
      }{
        (\fresh{j})
        \hsp
        \Alg \land 0 \leq j < n \land 0 \leq i < n \vd_{i,j}
          (H[is[es[i]]] = is[i] => es[i])
        \To{Inj=}~ \True
      }\textsc{(InjGe)}
      \vspace{0.5em}
    }{
      \Alg \vd \ixfn{\iot{i}{n}}{H[is[es[i]]] = is[i] => es[i]} \To{Inj} \True
    }\textsc{(InjF1)}
    \vspace{0.5em}
    \\
    \Env;\Alg \vd
    \ixfn{\iot{i}{n}}{
      \text{$\begin{array}{l}
      (c \land es[i] \in (-\infty,\infty) => es[i])
      \\
      \gdelim
      (\neg c \lor es[i] \notin (-\infty,\infty) => \infty)
      \end{array}$}
    }
    \leadsto
    \ixfn{\iot{i}{n}}{H[is[es[i]]] = is[i] => es[i]}
    % \ixfn{\iot{i}{n}}{c => es[i]}
    \\
    \Alg.FP(es') = (es, \Fun{i}{\overbrace{H[is[es[i]]] = is[i]}^{c}}, \_)
  }{
    \Env; \Alg \vd ((-\infty,\infty), es') \To{Inj} (\True, \Alg)
  }
  \textsc{(InjV2)}
\end{align*}
\end{footnotesize}
The query ({\tiny $\?$}) antecedent is sent to the equality solver with guide
$\textcolor{blue}{es[i]} = \textcolor{magenta}{es[j]}$, which is substituted
into the other terms before expanding the antecedent with transitive equalities:
\begin{small}
\[
  \textcolor{blue}{es[i]} = \textcolor{magenta}{es[j]}
    \land H[is[\textcolor{blue}{es[i]}]] = is[i]
    \land H[is[\textcolor{blue}{es[i]}]] = is[j]
    \land is[i] = is[j]
\]
\end{small}
\textsc{Eq.Inj} is then matches and simplifies the conjunct
$is[i] = is[j]$ to $i = j$ since $\Alg.Inj(is) = (-\infty,\infty)$.
% \lstinline[mathescape=true]{Inj is ($-\infty$,$\infty$)}
% with $-\infty \leq i \leq \infty$ and $-\infty \leq j \leq \infty$.
Finally, the expanded (now trivial) query is sent to the solver:
% \colorbox{black!6}
{
  \footnotesize
  $\Alg' \vd \textcolor{blue}{es[i]} = \textcolor{magenta}{es[j]} \land \dots \land i = j \? i=j$
}.
% The injectivity of the index function of $edges'$ is
% later used to verify the safety
% of another $\Scattertt$ operation (not shown).

\emph{Obviating dynamic bounds checks.}
We have verified that indexing is within bounds for a key computational kernel
from the sparse $k$-means benchmark in \cite{schenck2022ad}, shown in
\cref{fig:eval} (right).  For each indexing statement, a query is sent to the
solver that checks whether the indexing expression is within bounds of the array.
By gradually strengthening the preconditions, the user can let the compiler
guide them towards the weakest preconditions needed. For example, removing all
preconditions shown, the compiler reports:
\begin{lstlisting}
kmeans.fut:7:21-34: Unsafe indexing: pointers[row] (failed to show: True => 0 @$\leq$@ row).
\end{lstlisting}
which is rectified by adding precondition \lstinline[mathescape=true]{Range row (0,$\infty$)}.
Repeating this process until all errors are addressed yields
the preconditions shown in the figure.
We compile $k$-means to a CUDA program using the Futhark compiler~\cite{futhark-tuning}.
In \cref{tab:eval} (right), an average 2-times speed up is observed
on datasets movielens~\cite{movielens}, nytimes and scrna~\cite{nolet2022gpu}
(using parameters from~\cite{schenck2022ad})
by eliminating dynamic bounds checks.
In Futhark, dynamic bounds checking is handled by unstructured jumps
to the end of a CUDA kernel~\cite{henriksen2021bounds}, which likely
inhibits the {\tt nvcc} compiler from effectively optimizing 
instruction-level~parallelism.  A dynamic approach based on slicing
a safety predicate~\cite{slice-bounds-check} may be cheaper in this
case, but it will incur larger overheads to the simpler (common) cases.
% preventing further optimization of instruction-level~parallelism.
%
% FROM AD PAPER
% Sparse k-means performance measurements for three NLP work-
% loads. k = 10 for all datasets, with a fixed iteration count of 10 and a
% 32-bit representation. The movielens dataset uses data from the ML 20M
% dataset described in [34] with dimensions (139K, 131K) and a density of
% 0.11%. The nytimes and scrna datasets are the same as used in [35], with
% dimensions (300K, 102K) and (66K, 27K) and densities of 0.23% and
% 7.3%, respectively.
%

\emph{Segmented operations.}
We conclude our case studies with \texttt{partition2L},
a batched version of
%the \lstinline{indices}
%array in the body of
\texttt{partition2} (\cref{fig:running-egs}),
that applies to a jagged array with potentially empty rows.
The function takes as input an array \texttt{shp} of type
\texttt{[m]i64} whose elements denote the size of each
row in the jagged array and an array \texttt{csL}
of type \texttt{[n]bool}
where \texttt{n = sum shp},
and outputs an array of indices, that when scattered
will produce a partitioning of each row in the jagged
array according to \lstinline{csL}.
The full program (not shown here) makes use of
\lstinline{sgmSum}, \lstinline{mkSgmDescr}, and \lstinline{mkII} from
\cref{fig:running-egs}.
%
% ι  part2indicesL₄₇₉₁ =
%     {k₁₀₈₄₇₃ :: 0 .. m₄₇₄₃
%      i₁₅₀₁₈ :: ⊎k₁₀₈₄₇₃ [∑j₁₀₈₄₇₄∈(0 .. -1 + k₁₀₈₄₇₃) (shape₄₇₄₅[j₁₀₈₄₇₄]),
%                          ...,
%                          -1 + ∑j₁₀₈₄₇₄∈(0 .. k₁₀₈₄₇₃) (shape₄₇₄₅[j₁₀₈₄₇₄])]
%      forall i₁₅₀₁₈ . | csL₄₇₄₆[i₁₅₀₁₈] ⇒    ∑j₁₀₈₄₇₄∈(0 .. -1 + k₁₀₈₄₇₃) (shape₄₇₄₅[j₁₀₈₄₇₄]) + ∑j₁₀₈₄₇₄∈(∑j₁₀₈₄₇₅∈(0 .. -1 + k₁₀₈₄₇₃) (shape₄₇₄₅[j₁₀₈₄₇₅]) .. -1 + i₁₅₀₁₈) (csL₄₇₄₆[j₁₀₈₄₇₄])
%                      | ¬(csL₄₇₄₆[i₁₅₀₁₈]) ⇒    i₁₅₀₁₈ + ∑j₁₀₈₄₇₄∈(1 + i₁₅₀₁₈ .. -1 + ∑j₁₀₈₄₇₅∈(0 .. k₁₀₈₄₇₃) (shape₄₇₄₅[j₁₀₈₄₇₅])) (csL₄₇₄₆[j₁₀₈₄₇₄]),
The inferred index function of the indices used by the final \texttt{scatter},
denoted $ys$, intuitively describes the semantics of the program:
\vspace{-1ex}
{\footnotesize
\begin{align*}
% \ixfn{\cat{k=0}{e_m}\seg{j}{e_{k}}}{dd}
\union{k}{0}{m}{
  \ixfn{\seg{i}{\Sum{j}{0}{k-1}{shp[j]}}}{}}
    (cs[i] => \Sum{j}{0}{k-1}{shp[j]} + \Sum{j}{\Sum{j'}{0}{k-1}{shp[j']}}{i-1}{cs[j]})
    \gdelim (\neg cs[i] => i + \Sum{j}{i+1}{\Sum{j'}{0}{k}{shp[j']}-1}{cs[j]})
\end{align*}
}\vspace{-1ex}\\
For each segment $k = 0, \dots, m$, if $cs[i]$ is true,
then index $i$ maps to
the row offset, $\Sum{j}{0}{k-1}{shp[j]}$, plus
the number of trues in $cs[i]$ that come before
$i$ in that segment;
otherwise, index $i$ maps to $i$ plus the number of
trues that come after $i$ in that segment.
%
%We verify the semantics of
Verification of \texttt{partition2L} essentially comes down to verifying
$
InvFiltPart~ys~(\Fun{i}{true})~(m,k,\Sum{j}{0}{k-1}{shp[j]},\Fun{i}{cs[i]})
$,
which succeeds. %, where $ys$ is the output array.
% RESULTS
%
% kmeans-sparse:static-verification
% data/movielens.in:     127,943μs
% data/nytimes.in:       165,418μs
% data/scrna.in:         387,784μs
%
% kmeans-sparse:baseline
% data/movielens.in:     280,545μs
% data/nytimes.in:       315,594μs
% data/scrna.in:         861,014μs
%
% partition2:static-verification-plus-optimised
% [50000000]f32 [50000000]f32:         2614μs
% [100000000]f32 [100000000]f32:       5145μs
% [200000000]f32 [200000000]f32:      10557μs
%
% partition2:static-verification
% [50000000]f32 [50000000]f32:         2814μs
% [100000000]f32 [100000000]f32:       5565μs
% [200000000]f32 [200000000]f32:      11097μs
%
% partition2:baseline
% [50000000]f32 [50000000]f32:        12,357μs
% [100000000]f32 [100000000]f32:      38,947μs
% [200000000]f32 [200000000]f32:     135,742μs
\begin{figure}
\footnotesize
\begin{minipage}[t]{0.65\linewidth}
\begin{tabular}{lrrrrrrr}
\vspace{-0.5em}
                       &                        &               &              &              &                & \textsc{\% of}\\
\vspace{-0.5em}
                       & \textsc{Properties \&} &               &              &              & \textsc{Check} & \textsc{Compile}\\
\textsc{Program}       & \textsc{annotations}   & \textsc{Safe} & \textsc{\#S} & \textsc{\#A} & \textsc{time}  & \textsc{time}\\
\hline
\texttt{maxMatching}   & Range,~Equiv,~Inj          & $\checkmark$ &  6 & 14 & 0.7s & 29\% \\
\texttt{kmeans\_ker}   & Range                      & $\checkmark$ &  0 &  3 & 0.1s & 12\% \\
\texttt{partition2}    & Equiv,~FP                  & $\checkmark$ &  1 &  3 & 0.4s & 34\% \\
\texttt{partition3}    & Equiv,~FP                  & $\checkmark$ &  1 &  3 & 0.6s & 44\% \\
\texttt{partition2L}   & Range,~Equiv,~FP           & $\checkmark$ &  1 &  3 & 3.6s & 82\% \\
\texttt{filter}        & Equiv,~FP                  & $\checkmark$ &  1 &  3 & 0.3s & 27\% \\
\texttt{filter\_seg}   & Range,~Equiv               & $\checkmark$ &  1 &  3 & 1.6s & 68\% \\
\end{tabular}
\end{minipage}\begin{minipage}[t]{0.34\linewidth}
\begin{tabular}{lrrr}
\vspace{-0.4em}
\textsc{Program}     & \textsc{Base}   & \multicolumn{2}{c}{\textsc{Speedup}}  \\
\textsc{\& Data}     & \textsc{($ms$)} & \textsc{Static} & $+$\textsc{Opt}\\
\hline
\texttt{kmeans} & \\
\hsp movielens & 280 & $2.2\times$ \\
\hsp nytimes   & 315 & $1.9\times$ \\
\hsp scrna     & 861 & $2.2\times$ \\
\texttt{partition2} & \\
\hsp 50M       & 12  & $4.4\times$  & $4.7\times$\\
\hsp 100M      & 38  & $7.0\times$  & $7.5\times$\\
\hsp 200M      & 135 & $12.2\times$ & $12.8\times$\\
\end{tabular}
\end{minipage}
\vspace{-2ex}
\caption{
  Left:
  Summary of evaluated programs.
  IFP and FP abbreviate InvFiltPart and FiltPart, respectively.
  \textsc{Safe} reports whether all indexing and scatters are verified.
  \textsc{\#S} and \textsc{\#A} denote the number of scatters and
  annotations in the program.
  Check time is the time taken to infer index functions
  and prove properties (on a modern AMD CPU).
  Right: Performance results on an A100 GPU using Futhark's CUDA backend.
  Base and static denote, respectively, dynamic and static
  verification. +Opt removes initialisation of the scattered array.
}
\label{tab:eval}
\end{figure}

\emph{Additional benchmarks.}
We also verify and report on the following programs:
\texttt{partition2} and \texttt{partition3} which partition arrays
into two and three parts, respectively, and \texttt{filter}/\texttt{filter\_seg}
which filter a (segmented) array according to a predicate.
\Cref{tab:eval} summarizes the evaluation; compilation is timed using
Futhark's CUDA backend.  Check times increase when the index
functions are complex (\texttt{part2indicesL}) or there are many annotations
(\texttt{maxMatching}).  Still, most programs take less than one second to
check.

\cref{tab:eval} (right) shows the impact of static {\em{}vs.} dynamic 
verification of scatter in \texttt{partition2} compiled to CUDA
and run on arrays of random floats with 50, 100, and 200 million
elements.
The static version is further optimised by leaving the scattered
array uninitialized---safe because we prove that all locations
are overwritten (\textsc{Static}$+$\textsc{Opt}).
Speedup factors of 4--12 are observed:
dynamic verification of scatter breaks fusion opportunities,
increases the amount of irregular accesses to memory, and requires
the use of reduce-by-index~\cite{histo-sc20} to conform
with the work asymptotic. The latter uses atomic (min) accumulations
that thresh the L2 cache, thus exacerbating the overhead.
%dynamic verification exacerbates irregular accesses to memory
%in multiple buffers at the same time, which probably
%leads to threshing of the L2 cache.
%
% XXX
% Showing error message on proof that went wrong
% Maximal independent set
%

\section{Related work}
\label{sec:rel-work}

\emph{Liquid (Haskell).}
Liquid Types enhance Hindley-Milner type systems with
refinement types~\cite{rondon-liquid-2008,vazou-refinement-2014}, generating
verification conditions (VCs) in the QF-EUFLIA logic that SMT solvers check. In
Liquid Haskell, refinement reflection~\cite{vazou-refinement-2018} integrates
source functions into refinements and automates unfolding their definitions for
proofs, but still requires manual proofs (in a similar fashion to theorem
provers like Rocq or Agda) for non-trivial reasoning and mandates top-level
functions for each component. For example, verifying \texttt{partition2}
(without the final \texttt{scatter}) demands program restructuring and manual
proofs for properties like injectivity, which we still
could not establish. Verification of \texttt{partition3} is harder still and
\texttt{partition2L} adds another layer of complexity through the use of
segmented shapes and \texttt{scatter}, which is not an operation that can be
easily expressed in (Liquid) Haskell.
Our system, by contrast, uses index functions to automatically
infer dependencies and prove array properties without structural constraints,
freeing programmers from proof-writing, aiding domain experts, and separating
program and proof concerns.

\emph{$F^\star$ and Pulse.}  $F^\star$~\cite{Fstar} is a programming language and
proof assistant with dependent and refinement types, supporting effects and
combining SMT-based proof automation with interactive proof writing. It
automates term reasoning via reductions (similar to Liquid Haskell's rewriting)
and uses monadic effects to separate computation and proof. Still, it shares
many of Liquid Haskell’s limitations, requiring manual proofs for properties our
system handles automatically, despite a more ergonomic programming
discipline. Pulse~\cite{swamy2023proof}, embedded in $F^\star$, targets
concurrent programming with mutable state using Concurrent Separation
Logic~\cite{brookes2016concurrent} at the statement level to reason about
heaps. For instance, verifying an in-place version of \texttt{partition2} in the
context of quicksort checks the two buffers against a pivot and ensures
non-overlap, but this doesn’t extend to a data-parallel context that fisses the
computation into bulk operations and uses scatter to write all indices at once.

The reviewed approaches rely on powerful SMT solvers such as Z3~\cite{Z3}, which
uses e-graphs~\cite{Z3-uses-egraphs} to generate and test all equivalent
rewrites. Our rewrites of index and sum symbols (\cref{subsec:simplify}) are
effective, but are challenging to express in Z3. Our simplification strategy
uses directed rewrites, which are much cheaper computationally than e-graphs,
given that
\begin{enumerate*}
\item simplification must be performed after each symbol elimination,
\item each simplification may trigger many rewrites, and
\item the only possible objective function is whether the query succeeds.
\end{enumerate*}

\emph{Linear Array Logics.}  Dependent ML~\cite{XI_2007} and its extension
ATS~\cite{xi2017applied} restrict dependent values to a limited language to
ensure decidability. Dependent ML is parametric in this language, enabling
specialization for tasks like static array bounds checking via linear
constraints~\cite{DependentML-bounds-checking}. ATS further allows explicit
proof terms, though it requires intertwining proof and program despite their
syntactic separation. Other systems, such as the quantifier-free logic by Daca
et al.~\cite{array-folds-logic} for counting and partitioning, Bradley et al.'s
logic~\cite{whats-decidable-about-arrays} for index ranges and sortedness using
Presburger arithmetic, and Qube~\cite{Qube} for verifying array indexing and
shape matching, are restricted to linear indexing, which exclude operations %excluding indirect operations
like \lstinline{a[b[i]]}. In comparison, we target data-parallel programs with
non-linear indexing (e.g., gather/scatter/scan), which is the case of % in fact,
all benchmarks~evaluated~in~\cref{subsec:eq-with-inj}.

\emph{Dependece-Analyses on Arrays.} 
Our verification analysis takes inspiration from work in automatic
optimization of loop-based code. Related analyses have shown that
choosing a suitable representation for 
access patterns~\cite{set-congr-an,suif-dyn,HybAn} 
is key to scaling analysis interprocedurally, either 
statically~\cite{SUIF,PolyPluto2,CIVan}
or by combining static and 
dynamic~\cite{r-lrpd,HybAn,mem-ref-mon}
to cover challenging non-affine code instances.
Importantly, dynamic analyses use complex transformations such as
program slicing and hoisting to extract and test at runtime sufficient
conditions for statically irreducible queries. These often boil down
to establishing array properties such as 
permutation~\cite{it-data-reorder-2,it-data-reorder-1}, 
injectivity~\cite{HybAn,r-lrpd},
monotonicity~\cite{mem-ref-mon,CIVan}.
% of arrays used in sparse computations. 
By establishing array properties early, our approach could in 
principle simplify dependence analyses and eliminate expensive
runtime overheads. 

In hopeless cases that require some dynamic
verification,\footnote{One example is the Brownian Bridge component of
the option pricing describe in~\cite{fhpc-pricing}, which uses three 
indirect arrays.} property specification still allows efficient
code generation and placement of the inspector code (e.g., avoiding 
program slicing and hoisting).
Finally, reminded by work on parametric polymorphism~\cite{mapal_synasc,pp-sca-oopsla},
specification of array properties should be supported across multi-language
components, which may separate the definition of an array from its use; otherwise 
optimization of such codes may require suboptimal speculation~\cite{jaycos:DTLS}. 

\emph{Scheduling Languages \& Verfication of Compiler Transformations.} 
Work on verifying the specification, code-transformations and the resulting
low-level code of scheduling DSLs such as Halide~\cite{Halide} include
improvements to its term rewriting system~\cite{Halide-Verif-Term-Rewrite},
%improving the term rewriting system of the expression
%simplifier~\cite{Halide-Verif-Term-Rewrite},
end-to-end translation validation of {\em affine}
specifications~\cite{Halide-poly-valid}, and HaliVer~\cite{HaliVer}, which aims
to
%uses deductive program verification to
verify properties of (1) the specification, which utilize {\em linear} indexing,
and (2) of the low-level generated code, which employs permission-based
separation logic~\cite{permission-sep-logic} to verify, e.g., memory safety.
%The former verifies properties that use {\em linear} indexing,
%including counting.
%The latter, uses permission-based separation logic~\cite{permission-sep-logic}
%and VerCors~\cite{VerCors} to verify memory safety and functional correctness.
%
Finally, bounded translation validation tools such as Alive2~\cite{Alive,Alive2}
are used to verify code transformations of LLVM.  These directions are
complementary to our work.

%%%%%%%%%%%%%%%%%%%%%%%%%%%%%%%%%%%%%
%%% Conclusions
%%%%%%%%%%%%%%%%%%%%%%%%%%%%%%%%%%%%%

\section{Conclusions}
We presented a framework for inferring and verifying properties of integral
arrays in the context of a pure data-parallel array language. Our solution
supports a small but powerful set of properties---namely range, equivalence,
monotonicity, injectivity, bijectivity, and filtering/partitioning---that are
easy to use and expose a rich compositional algebra to the compiler.  This is
used to scale (automate) the analysis (restricting user intervention to
strategic places) and to optimize the program.

Our evaluation demonstrates that our framework is capable of verifying
challenging code patterns from graph algorithms and from flattening irregular
nested parallelism (e.g., the batch application of two-way partitioning to
jagged arrays in flat form).
It also shows that eliminating dynamic checks and redundant initializations
results in significant GPU speedups, and that our design facilitates
debugging---e.g., by user inspection of the context and index function(s).

%\clearpage

%%%%%%%%%%%%%%%%%%%%%%%%%%%%%%%%%%%%%
%%% End of 23 pages paper
%%%%%%%%%%%%%%%%%%%%%%%%%%%%%%%%%%%%%

\bibliographystyle{ACM-Reference-Format}
\bibliography{main.bib}

\end{document}